\documentclass[reqno,12pt,a4paper]{amsart}
\usepackage{graphicx}
\usepackage{psfrag}

\usepackage{mathrsfs}
\usepackage{amssymb}
\usepackage{amsfonts}
\usepackage{latexsym}
\usepackage{amsmath}
\usepackage{amsthm}

\newcommand{\er}[1]{\textrm{(\ref{#1})}}
\def\lb{\label}

\theoremstyle{plain}

\newtheorem{theorem}{\bf Theorem}[section]
\newtheorem{lemma}[theorem]{\bf Lemma}
\newtheorem{proposition}[theorem]{\bf Proposition}

\theoremstyle{remark}

\renewcommand{\a}{\alpha}           \newcommand{\cA}{\mathcal{A}}

\renewcommand{\b}{\beta}            \newcommand{\cB}{\mathcal{B}}

\newcommand{\g}{\gamma}           \newcommand{\cC}{\mathcal{C}}

\renewcommand{\d}{\delta}           \newcommand{\cE}{\mathcal{E}}

\newcommand{\ve}{\varepsilon}     \newcommand{\cG}{\mathcal{G}}

\newcommand{\e}{\eta}

\renewcommand{\k}{\kappa}           

\renewcommand{\l}{\lambda}          \newcommand{\cM}{\mathcal{M}}

\newcommand{\m}{\mu}

\newcommand{\s}{\sigma}

\newcommand{\vp}{\varphi}         

\renewcommand{\c}{\chi}             
 
\newcommand{\p}{\psi}             
 
             \newcommand{\cZ}{\mathcal{Z}}
      
\renewcommand{\o}{\omega}
\renewcommand{\O}{\Omega}
\newcommand{\x}{\xi}

\newcommand{\vk}{\varkappa}

  \def\mA{{\mathscr A}}
 \def\mB{{\mathscr B}}

  \def\mH{{\mathscr H}}

\def\Z{\mathbb{Z}}
\def\R{\mathbb{R}}
\def\C{\mathbb{C}}

\def\N{\mathbb{N}}

\def\qqq{\qquad}
\def\qq{\quad}

\let\ge\geqslant
\let\le\leqslant

\newcommand{\ca}{\begin{cases}}
\newcommand{\ac}{\end{cases}}
\newcommand{\ma}{\begin{pmatrix}}
\newcommand{\am}{\end{pmatrix}}

\def\lt{\biggl}
\def\rt{\biggr}

\renewcommand{\[}{\begin{equation}}
\renewcommand{\]}{\end{equation}}

\def\wt{\widetilde}
\def\pa{\partial}
\def\sm{\setminus}

\def\no{\noindent}
\def\ol{\overline}
\def\iy{\infty}
\def\ev{\equiv}
\def\/{\over}
\def\ts{\times}

\def\Re{\mathop{\rm Re}\nolimits}

\def\Im{\mathop{\rm Im}\nolimits}

\def\Tr{\mathop{\rm Tr}\nolimits}

\def\diag{\mathop{\rm diag}\nolimits}

\def\BBox{\hspace{1mm}\vrule height6pt width5.5pt depth0pt \hspace{6pt}}
\def\wh{\widehat}
\def\as{\text{as}}
\def\where{\text{where}}
\def\1{1\!\!1}

\def\bu{\bullet}

\begin{document}

\title[Inverse problems and eigenvalue asymptotics
] {Inverse problems and sharp eigenvalue asymptotics for
Euler-Bernoulli operators}

\date{\today}
\author[Andrey Badanin]{Andrey Badanin}
\address{Mathematical Physics Department, Faculty of Physics,
Ulianovskaya 2,
St. Petersburg State University, St. Petersburg, 198904,
 Russia,}
\author[Evgeny Korotyaev]{Evgeny Korotyaev}
\address{
an.badanin@gmail.com
\newline
korotyaev@gmail.com}

\subjclass{47E05, 34L20}
\keywords{ Euler-Bernoulli operator, fourth
order operator, inverse problem, eigenvalue asymptotics} \maketitle

\centerline
{\it Dedicated to the memory of Professor Vladimir Shelkovich,
1949-2013}

\begin{abstract}
We consider Euler-Bernoulli operators with real coefficients on the
unit interval. We prove the following results:

 i)  Ambarzumyan type theorem about the
inverse problems for the Euler-Bernoulli operator.

ii) The sharp asymptotics of eigenvalues
for the Euler-Bernoulli operator
when its coefficients converge to the constant function.

iii) The sharp eigenvalue asymptotics both for the Euler-Bernoulli
operator and  fourth order operators (with complex coefficients) on
the unit interval at high energy.

\end{abstract}

\section {Introduction and main results}
\setcounter{equation}{0}

\subsection{Euler-Bernoulli operators}
We consider the Euler-Bernoulli operator $\cE$ given by
\[
\lb{defcE}
\cE u={1\/b}(au'')''+ Q u, \qqq
\]
acting on $L^2((0,1),b(x)dx)$ with the boundary conditions
\[
\lb{ebdc}
\begin{aligned}
u=0,\qqq \text{and} \qq u''+2(\a+\b) u'=0\qq
\text{at}\qq x=0 \qq \text{and} \qq x=1,
\end{aligned}
\]
where $a,b$ are positive coefficients given by
\[
\lb{abxe}
a(x)=e^{4\int_0^x\a(s)ds}>0,\qqq
b(x)=b(0)e^{4\int_0^x\b(s)ds}>0,
\]
without loss of generality we assume $a(0)=1$.
We assume that the functions $\a,\b,  Q$ are real and satisfy:
\[
\lb{abQsp} (\a,\b, Q)\in \mH_3\ts\mH_3\ts\mH_0,
\]
where $\mH_m$ is the Sobolev space defined  by
\[
\lb{mHm} \mH_m=\Big\{f\in L^1(0,1): f^{(m)}\in L^1(0,1)\Big\},\qqq
 m\ge 0.
\]

The Euler-Bernoulli operator is a specific form of a fourth order
operator. It describes the bending vibrations of thin elastic beams,
$a$ is the rigidity and $b$ is the density of the beam,
see \cite{Gl1}. The boundary
conditions \er{ebdc} mean that the ends of the beam are restrained
by some special rotational spring devices.
If $(\a+\b)|_0=(\a+\b)|_1=0$ (for example,
if the coefficients $a,b$ are constant near the ends), then
\er{ebdc} implies the boundary conditions for the pinned-pinned
beam:
\[
\lb{pinc}
u(0)=u(1)=u''(0)=u''(1)=0.
\]

In order to get $(\pi n)^4$ in the first term in asymptotics
\er{EBasDir}, we add a following normalizations
\[
\lb{sigmv} \int_0^1\Big({b\/a}\Big)^{1\/4}dx=1,
\]
without loss of generality. The operator  $\cE$ is self-adjoint and
its spectrum   is real, purely discrete and consists of eigenvalues
$\l_n,n\in\N$, of multiplicity $\le 2$ labeled by
$$
\l_1\le\l_2\le\l_3\le...,
$$
counted with multiplicity. In the case of a uniform beam (i.e.,
$a,b=1$) conditions \er{ebdc} takes the form \er{pinc} and the
corresponding eigenvalues have the form  $\l_n=(\pi n)^4,n\in\N$.

Recall the following famous result of Ambarzumyan
(see \cite[Ch VI]{LeS}):

\no {\it Let $\l_0<\l_1<...$ be eigenvalues of the  problem
$$
-y''+q(x)y=\l y,\qqq x\in[0,1],\qqq
y'(0)=y'(1)=0,
$$
where $q$ is a real continuous function. Then  $\l_n=(\pi n)^2$ for
all $n=0,1,2,...$, iff  $q=0$. }

Note that,
in general,  spectrum  of the second order operator does not
determine the potential, i.e., Ambarzumyan's theorem is not valid
for other boundary condition.
The results, similar to Ambarzumyan's theorem,
for the weighted second order operator
${1\/b}u''$ are well known, see discussions in Section 5.

In order to formulate  Ambarzumyan type results about the inverse
problems for the Euler-Bernoulli operator we define the following
functions:
\[
\lb{kapx}
\k={3\a+5\b\/2\x},\qqq \x=\Big({b\/a}\Big)^{1\/4}.
\]

\begin{theorem}
\lb{cor1}
Let $Q=0$ and let
the real $(\a,\b)\in \mH_3\ts\mH_3$ satisfy the
conditions \er{sigmv} and $\k(0)=\k(1)$. Then the eigenvalues
$\l_n=(\pi n)^4$ for all $n\ge 1$ iff $a=b=1$.

\end{theorem}

\no{\bf Remark.} In contrast to second order operators
$-y''+qy$, we conjecture that this result  may be extended onto
Euler-Bernoulli operators with some other boundary conditions, see
details in Section 5 after the proof of Theorem.

\medskip
%%%%%%%%%%%%%%%%%%%%

The proof of Theorem \ref{cor1} is based on eigenvalue asymptotics
for the operator $\cE$. In order to determine this asymptotics we
introduce a constant $\p_0$ and a function $\vk$ by
\[
\lb{psi} \p_0=\k(1)-\k(0)+\int_0^1{\vk(x)\/\x(x)}dx, \qqq
\vk={5\a^2+5\b^2+6\a\b\/4}\ge {\a^2+\b^2\/2}\ge 0.
\]

\begin{theorem}
\lb{evasEuB}
Let $(\a,\b, Q)\in \mH_3\ts\mH_3\ts\mH_0$ be real and
be normalized by \er{sigmv}. Then the eigenvalues $\l_n$ of the
operator $\cE$ satisfy
\[
\lb{EBasDir}
\l_n=(\pi n)^4+2(\pi n)^2\p_0+\p_1-\g_{n}+{o(1)\/n}
\qqq\as\qq n\to\iy,
\]
uniformly on  bounded subsets of $\mH_3\ts\mH_3\ts\mH_0$, where
$\p_1$ is a constant given by \er{psi1}, and
\[
\lb{ebg} \g_n=\int_0^1\Big( Q(x)\x(x)
+{\a'''(x)-\b'''(x)\/4\x^3(x)}\Big)
\cos\Big(2\pi n\int_0^x\x(s)ds\Big)dx.
\]

\end{theorem}

\no {\bf Remark.} 1) If we introduce a new variable
$t=t(x)=\int_0^x\x(s)ds$ in the integral in \er{ebg}, then $\g_n$
become Fourier coefficients of the function $
Q+{b\/4a}(\a'''-\b''')$.

2) Jian-jun, Kui and  Da-jun \cite{JKD} announced the first two
terms in the eigenvalue asymptotics for
the Euler-Bernoulli equation with smooth coefficients on the finite
interval for other boundary conditions, see more in Section 5.

3)  Let $\a,\b$ satisfy $\k(1)=\k(0)$ (e.g., $a,b$ are periodic).
Then the definition of $\p_0$ given by \er{psi}  shows that the
constant $\p_0=\p_0(\a,\b)$ satisfies
$$
\p_0(\a,\b)=\p_0(-\b,-\a)=\int_0^1{\vk(x)\/\x(x)}dx\ge0.
$$
Moreover, this yields that $\p_0(\a,\b)=0$ iff $\a=\b=0$.
Thus we obtain that any periodic perturbation of the coefficients
$a,b$ moves strongly  all large eigenvalues to the right. Moreover,
eigenvalue asymptotics for the operators $\cE(a,b)$ and
$\cE({1\/b},{1\/a})$ coincide up to $O(1)$.

4) The proof of Theorem \ref{evasEuB} is based on the sharp
eigenvalue asymptotics for the operator $H=\pa^4+2\pa p\pa +q$ on
$L^2(0,1)$ from Theorem \ref{4g.thasdir}. The unitary
Barcilon-Gottlieb transformation \cite{B2}, \cite{Go1} (i.e., the
Liouville type transformation for a fourth order operator) reduces
the Euler-Bernoulli operator to the operator $\pa^4+2\pa p\pa +q$
with some specific  $p,q$. We give the derivation of the
Barcilon-Gottlieb transformation in Section 6.

\subsection{Euler-Bernoulli operators
with near constant coefficients.} Theorem \ref{evasEuB} shows that
any periodic perturbation of the coefficients $a,b$ shifts all large
eigenvalues to the right. In order to understand the situation for
the finite eigenvalues we consider  the operator $\cE_\ve $  given
by
\[
\lb{cEsm}
\cE_\ve u={1\/b^\ve}(c_\ve a^\ve u'')''+\ve Q u, \qqq
u(0)=u(1)=u''(0)=u''(1)=0,
\]
where  real $\ve\to 0$, the functions $a, b$ have the form
\er{abxe}, the constant $c_\ve>0$ is defined by
\[
\lb{cvesm}
c_\ve^{1\/4}=\int_0^1\Big({b(x)\/a(x)}\Big)^{\ve\/4}dx,
\]
and $\a,\b,Q\in L^1(0,1)$ are  real.
Due to \er{cvesm} the coefficients $c_\ve a^\ve$,
$b^\ve$  are normalized by the identity $ \int_0^1({b^\ve\/c_\ve
a^\ve})^{1\/4}dx=1$ for all $\ve\in\R$, similar to \er{sigmv}. Let
$\l_n(\ve),n\in\N$, be eigenvalues of the operator $\cE_\ve$ labeled
by $\l_1(\ve)\le \l_2(\ve)\le \l_3(\ve)\le ...$ counted with
multiplicity.

We define Fourier coefficients  $\wh f_0, \wh f_{cn}, \wh f_{sn},
n\in\N$ of a function $f$ by
\[
\lb{Fc} \wh f_0=\int_0^1f(t)dt,\qqq \wh f_{cn}=\int_0^1
f(t)\cos(2\pi nt)dt,\qq \wh f_{sn}=\int_0^1 f(t)\sin(2\pi nt)dt.
\]

\begin{theorem}
\lb{Thsm}
Let $\a,\b,Q\in L^1(0,1)$. Then each eigenvalue
$\l_n(\ve), n\ge 1$, of the operator
$\cE_\ve$ is analytic in $\{\ve\in\C:|\ve|<\ve_1\}$
for some $\ve_1>0$
and satisfies
\[
\lb{aslnsm}
\l_n(\ve)=(\pi n)^4+2\ve(\pi n)^3(\wh\a_{sn}-\wh\b_{sn})
+\ve(\wh Q_0-\wh Q_{cn}) +O(\ve^2)
\]
as $\ve\to 0$, uniformly in $\a,\b,Q$
on bounded subsets of $L^1(0,1)$.
\end{theorem}

\no{\bf Remark.}
1) Asymptotics \er{aslnsm} shows that each perturbed
eigenvalue $\l_n(\ve)$ remains closed to the unperturbed one
$\l_n(0)=(\pi n)^4$  under the small perturbations. They can move to
left or to right. In particular, if $Q=0$, then \er{aslnsm} gives
\[
\lb{derevsc}
\l_n'(0)=2(\pi n)^3(\wh\a_{sn}-\wh\b_{sn}).
\]
 If $\wh\a_{sn}>\wh\b_{sn}$, we obtain
$\l_n(\ve)>\l_n(0)$. If $\wh\a_{sn}<\wh\b_{sn}$, then  we obtain
$\l_n(\ve)<\l_n(0)$.

2) Badanin-Korotyaev \cite{BK1}, \cite{BK2} considered
fourth order operators with small coefficients
in the periodic case.

\medskip

Barcilon \cite{B1} considered
a boundary value problem
\[
\lb{Baeq}
\big((1+\ve\a)u''\big)''=\l (1+\ve\b)u,
\qqq
u(0)=u'(0)=u''(1)=(au'')'(1)=0.
\]
He solved a problem of reconstruction of coefficients $\a,\b$
by the first term of the perturbation series
for eigenvalues, as $\ve\to 0$.
Describe briefly the results of Barcilon.
Let $\m_n(\ve)$ be eigenvalues of the problem \er{Baeq}.
Barcilon  proved that the coefficients $\a$ and $\b$
cannot be uniquely determined by the sequence $\m_n'(0),n\in\N$.
Moreover, he showed that it is sufficient to know
the spectra of three Euler-Bernoulli operators
with different boundary conditions
in order to uniquely determine both $\a$ and $\b$.
Asymptotics \er{aslnsm} gives a solution of the Barcilon
inverse problems in our case.

\medskip

\no {\bf Example.}
Consider our Euler-Bernoulli operator $\cE_\ve$ as $\ve\to 0$, which
corresponds to the case of $a,b$ closed to one.
We show that if we know
the coefficient $\a$ and we know, in addition, that $\b$
is odd, then $\b$ can be uniquely recovered by the sequence
$\l_n'(0)$, given by \er{derevsc}.

Assume that $\a\in L^1(0,1),Q=0$ and
for some unknown
$$
\b\in L_{odd}^1(0,1)=\{f\in L^1(0,1):f(x)=f(1-x),x\in(0,1)\}
$$
we have the sequence $\l_n'(0),n\in\N$, of derivatives
of eigenvalues of the operator $\cE_\ve$ at $\ve=0$.
Identity \er{derevsc} gives
$$
\sum_{n=1}^\iy(\wh\a_{sn}-\wh\b_{sn})\sin 2\pi nx=
{1\/2}\sum_{n=1}^\iy{\l_n'(0)\/(\pi n)^3}\sin 2\pi nx,\qqq
x\in(0,1).
$$
Then $\b$ is uniquely determined by
\[
\lb{fcb}
\b(x)={1\/2}\Big(\a(x)-\a(-x)-
\sum_{n=1}^\iy{\l_n'(0)\/(\pi n)^3}\sin 2\pi nx\Big),\qqq
x\in(0,1).
\]

\medskip

\no {\bf Remark.}
1) The similar arguments show that the function
$\a\in L_{odd}^1(0,1)$ can be determined by
$\b\in L^1(0,1)$ and $\l_n'(0),n\in\N$.

2) Solution of an inverse spectral problem for
a second order operator on the unit interval
under the Dirichlet boundary condition for even (and generic)
potentials
see in P\"oschel-Trubowitz \cite{PT}, the case of
a weighted operator was considered by
Coleman-McLaughlin \cite{CM}.

\subsection{Fourth order operators.}
 We consider an operator $H$ on $L^2(0,1)$ given by
\[
\lb{op} Hy= y''''+2(py')'+qy,
\]
under  the boundary conditions
\[
\lb{4g.dc}
y(0)=y''(0)=y(1)=y''(1)=0.
\]
In our paper, in general, the operator $H$ is non-selfadjoint, since
we assume that the functions $p,q\in L^1(0,1)$ are complex
(including real).

It is well known (see, e.g., \cite[Ch.~I.2]{Na}) that  the spectrum  of
the operator $H$ is purely discrete and consists of eigenvalues
$\l_n,n\in\N$, labeled by
$$
|\l_1|\le|\l_2|\le|\l_3|\le...,
$$
counted with multiplicities. Furthermore, the following eigenvalue
asymptotics hold true:
$$
\l_n=(\pi n)^4+O(n^{2})\qqq\as\qq n\to\iy,
$$
see, e.g., \cite[Ch.~I.4]{Na}. Note that in the case of complex
coefficients we mean the "algebraic" multiplicity, i.e.,  a total
dimension of the root subspace corresponding to the eigenvalue. This
"algebraic" multiplicity may be any integer, whereas the "geometric"
multiplicity, i.e. the dimension of the eigen-subspace, is $\le 2$.
In our paper we determine the eigenvalue asymptotics for the
operator $H$.

Note that in the case of real coefficients any self-adjoint fourth
order operator may be written in the form \er{op}. Moreover, if the
functions $p,q$ are real, then the eigenvalues $\l_n,n\in\N$, are
real, satisfy $\l_1\le\l_2\le\l_3\le... $, the "algebraic"
multiplicity is $\le 2$ and coincides with the "geometric"
multiplicity.

The most sharp eigenvalue asymptotics for the  operator $H$
at the present time was determined by
Caudill-Perry-Schueller \cite{CPS}.
They proved that in the case of the real coefficients
$p,q\in L^1(0,1)$ the eigenvalues $\l_n$ satisfy
\[
\lb{4g.T2-2}
\l_n=(\pi n)^4 -2(\pi n)^2\big(\wh p_0+ \wh
p_{cn}\big)+O(n^{1+\ve})
\]
as $n\to\iy$ for any $\ve>0$. Note that there are no coefficients
$\wh q_{cn}$ in the leading term of asymptotics \er{4g.T2-2}, since
$p\in L^1(0,1)$. In order to get $q$ in  leading  term of eigenvalue
asymptotics we need an additional condition $p''\in L^1(0,1)$.

In Theorem \ref{4g.thasdir} we determine the sharp eigenvalue
asymptotics of the operator $H$ for complex coefficients  $p, p'',
q\in L^1(0,1)$. In fact, the asymptotics is expressed
 in terms of the Fourier
coefficients of a function $ V=q-{p''\/2}$. We formulate our main
results about the eigenvalue asymptotics for fourth order operators.

\begin{theorem}
\lb{4g.thasdir} Let $(p,q)\in \mH_{2+j}\ts\mH_j$, where  $j\in \{0,
1\}$, and let $V=q-{p''\/2}$.
 Then the eigenvalues $\l_n$  of the operator $H$ satisfy the
asymptotics
\[
\lb{4g.asDir} \l_n=(\pi n)^4-2(\pi n)^2 \wh p_0
-{1\/2}\int_0^1(p^2(t)-\wh p_0^2)dt +\wh V_0- \wh
V_{cn}+{\ve_n\/n^{1+j}},\qqq \ve_n=\ca o(1), j=0\\
O(1), j=1\ac
\]
as $ n\to\iy$, uniformly on bounded subsets of $\mH_{2+j}\ts\mH_j$,
where $\wh p_0,\wh V_0,\wh V_{cn}$ are defined by \er{Fc}.

\end{theorem}

\no {\bf Remark.} 1) Asymptotics \er{4g.asDir} were used to prove
trace formulas for fourth order operators on both the circle
\cite{BK5} and the unit interval \cite{BK6}.

2) The proof of asymptotics \er{4g.T2-2} in \cite{CPS} is based on
analysis of both the free and perturbed resolvents. Our  approach is
different and is  based on analysis of a determinant defined by
\er{defD}, expressed in terms of the fundamental solutions, which
are entire in $\l$. All eigenvalues $\l_n, n\ge 1$ are zeros of this
determinant. Using the sharp asymptotics of the fundamental
solutions from \cite{BK4} we determine asymptotics of the
determinant (the corresponding proof is rather technical, see
Section 7). Analysis of this asymptotics provides the sharp
asymptotics of $\l_n$.

3) In some papers there are misprints and mistakes in eigenvalue
asymptotics for the second and fourth order operators. We discuss
some of them in Section 4.

\subsection{Historical review.}

\no $\bu$ {\it Second order operators.} There are a lot of results
about the eigenvalue asymptotic for second order operators, see the
book of Levitan-Sargsyan \cite{LeS}, the review of Fulton-Pruess
\cite{FP} and  the references therein. Atkinson-Mingarelli \cite{AM}
obtained the eigenvalue asymptotics $\l_n=(\pi n)^2(1+o(1))$ as
$n\to\iy$, for the weighted second order operator $b^{-1}u''$, where
the coefficient $b$ satisfies $b\in L^1(0,1)$ and $ b^{-1}\in
L^\iy(0,1)$. Korotyaev \cite{K} determined the sharp eigenvalue
asymptotics in terms of the Fourier coefficients of $b'/b$, where
the coefficient $b$ satisfies $ b, b^{-1}\in L^\iy(0,1)$ and $b'\in
L^2(0,1)$. Asymptotics for the case of smooth coefficients were
determined by asymptotics for the Schr\"odinger operator using
Liouville transformation, see Fulton-Pruess \cite{FP}.

Sufficiently sharp eigenvalue asymptotics for the Schr\"odinger
operator with a matrix potential on the finite interval   were
determined  by Chelkak-Korotyaev for both the Dirichlet boundary
conditions \cite{CK1}  and for the periodic boundary conditions
\cite{CK2}. Now we describe the difference between the asymptotics
for the systems of second order equations and the scalar higher
order differential equations. The unperturbed fundamental matrix for
the second order operators (even with the matrix-valued
coefficients) have the entries $\cos\sqrt\l t$ and $\sin\sqrt\l t$.
Then all entries of the perturbed fundamental matrix are bounded as
$\l\to+\iy$ and their asymptotics can be determined by the standard
iteration procedure. But in the higher order case we meet some
additional difficulties, which are quite absent for second order
operators. For example, the unperturbed fundamental matrix for the
fourth order operators has both the bounded entries $\cos\l^{1\/4}t$
and $\sin\l^{1\/4}t$ and the unbounded entries $\cosh\l^{1\/4}t$ and
$\sinh\l^{1\/4}t$  as $\l\to+\iy$. Therefore, the standard
iterations don't give asymptotics of the perturbed fundamental
matrix for the fourth (and higher) order operators. Roughly speaking
the determining of eigenvalue asymptotics for higher order operators
has the standard  difficulties of analysis of the systems plus
additional ones associated with increasing parts of the fundamental
matrix.

$\bu$ {\it Fourth  order operators.}
 We discuss some results
for the fourth order operators. Such operators arise in many
physical models, see, e.g., the book \cite{PeT} and references
therein. Numerous results about the regular and singular boundary
value problems for higher order operators are expounded in the books
of Atkinson \cite{At}, Naimark \cite{Na}.

Eigenvalue asymptotics for fourth and higher order operators on the
unit interval are much less investigated than for the second order
operators. The first term is well known, see Naimark \cite{Na}. The
second term is known due to Caudill-Perry-Schueller \cite{CPS}. Note
that Badanin-Korotyaev  \cite{BK3} determined the corresponding term
in the eigenvalue asymptotics for general case of the $2n$ order
operators on the circle (for the case $n=2$ see \cite{BK2},
\cite{BK4}). It is more difficult to obtain the next term. The
simple case $\pa^{2n}+q$ was considered by Akhmerova \cite{Ah},
Badanin-Korotyaev \cite{BK1}, Mikhailets-Molyboga \cite{MM}.

Many papers are devoted to the inverse spectral problems for fourth
order operators. Barcilon \cite{B1} considered the inverse spectral
problem for the fourth order operators on the interval $[0,1]$ by
three spectra. McLaughlin \cite{McL1} studied the inverse spectral
problems by the spectrum and the norming  constants.
Caudill-Perry-Schueller \cite{CPS} described iso-spectral potentials
for our fourth order operators $H$. Hoppe-Laptev-\"Ostensson
\cite{HLO} considered the inverse scattering problem for the fourth
order operator on the real line in the case of rapidly decaying
$p,q$ at infinity. Yurko \cite[Ch 2]{Yu} recovered coefficients of a
fourth order operator on the unit interval by its Weyl matrix.

$\bu $ {\it The Euler-Bernoulli equation}. Jian-jun, Kui and Da-jun
\cite{JKD} determined two terms in the formal asymptotic eigenvalue
expansion for the Euler-Bernoulli operator on the unit interval.
Many papers are devoted to the inverse spectral problems for the
Euler-Bernoulli equation: Barcilon \cite{B2}, Gladwell \cite{Gl2},
Gottlieb \cite{Go1}, McLaughlin \cite{McL1}, see also the book of
Gladwell \cite{Gl1} and references therein. Papanicolaou \cite{P}
considered the inverse problem for the the Euler-Bernoulli on the
line in the case of periodic $a,b>0$. Moreover, there is enormous
physical and engineering literature on the Euler-Bernoulli equation,
here we mention only some papers related to our subject: Ghanbari
\cite{Gh}, Gladwell, England and Wang \cite{GEW}, Gladwell and
Morassi \cite{GlM}, Gottlieb \cite{Go2}, Guo \cite{Guo}, Chang and
Guo \cite{ChG}, Kambampati, Ganguli and Mani \cite{KGM}, Kawano
\cite{Ka}, Lesnic \cite{Ls}, Soh \cite{Soh}, Sundaram and
Ananthasuresh \cite{SuA}.

\subsection{The plan of our paper}
The plan of the paper is as follows. In Section 2 we prove the main
theorems. In Section 3 we discuss examples and remarks about
eigenvalue asymptotics for both second order and fourth order
operators. In Section 4 we consider the Euler-Bernoulli operators
with near constant coefficients. In Section 5 we consider the
Barcilon-Gottlieb transform, i.e., the unitary transformation between
the operator $H$ and the Euler-Bernoulli operator. In Section 6 we
determine asymptotics of the determinant $D$ whose zeros constitute
the spectrum of $H$.

\section{Proof of main theorems}
\setcounter{equation}{0}

\subsection{Eigenvalue asymptotics for the Euler-Bernoulli operator}
Consider the operator $\cE$ given by \er{defcE}, \er{ebdc}.
Introduce the new variable $t\in[0,1]$ by
\[
\lb{tx}
t(x)=\int_0^x\x(s)ds,\qqq\forall\qq x\in[0,1], \qq \where
 \qq \x=\Big({b\/a}\Big)^{1\/4}>0.
\]
Let $x(t)$ be the inverse
function for $t(x), x\in [0,1]$. Introduce the unitary transformation
$U:L^2((0,1),b(x)dx) \to L^2((0,1),dt)$ by
\[
\lb{defU}
(Uu(x))(t)=\rho(x(t))u(x(t))\qqq\forall\qq t\in[0,1],
\]
where
\[
\lb{deff}
\rho=a^{1\/8}b^{3\/8}>0.
\]
We show that the operators $\cE$ and $H$ are unitarily
equivalent, see the proof in Section 6.

\begin{lemma}
\lb{lmEuB} Let an operator $\cE$ be defined by \er{defcE},
\er{ebdc}, where $(\a,\b,Q)\in\mH_3\ts\mH_3\ts\mH_0$. Let the
operator $H$ be given by \er{op}, \er{4g.dc}, where the functions
$p(t),q(t), t\in [0,1]$ have the forms
\[
\lb{peb}
p=\vp-\s^2-2\s_t',
\]
\[
\lb{qeb}
q=-\s_{ttt}'''+(\s_{t}')^2+{4\/3}(\s^3)_{t}'
+\s^4-2(\vp\s)_{t}'-2\vp\s^2+\upsilon,
\]
and
\[
\lb{vp5}
\vp(t)={1\/2\x(x)}\Big({\e_-(x)\/\x(x)}\Big)_x'
+{\e(x)\/\x^2(x)},\qq
\s(t)={s(x)\/\x(x)},\qq
\upsilon(t)= Q(x)\qq\text{at}\ \ x=x(t),
\]
\[
\lb{sphi} s={\a+3\b\/2},\qqq \e=\e_+\e_-,\qqq \e_\pm=\b\pm\a.
\]
Let the unitary operator $U$ be defined by \er{defU}.
 Then the operators $\cE$ and $H$ are unitarily equivalent and
satisfy:
\[
\lb{eqEH} \cE=U^{-1}HU.
\]

\end{lemma}

We need the following identity.

\begin{lemma}
\lb{LmmAmB} Let $\a,\b\in\mH_3$. Then the function $p$ has the form
\[
\lb{idp1}
p=-{\vk\/\x^2}-\k_t',
\]
where $\vk$  and  $\k$ are given by \er{psi} and by \er{kapx}
respectively.
\end{lemma}

\no {\bf Proof.}
Identities  \er{vp5} and $s^2-\e=\vk$ imply
\[
\lb{ssq-vp}
\s^2-\vp={s^2-\e\/\x^2}-{1\/2}\Big({\e_-\/\x}\Big)_t'
={\vk\/\x^2}-{1\/2}\Big({\e_-\/\x}\Big)_t'.
\]
Substituting \er{ssq-vp} into \er{peb} and   using the identity
${1\/2\x}(\e_--4s)=-\kappa$, we obtain
$$
p=-{\vk\/\x^2}+\Big({\e_-\/2\x}-2\s\Big)_t'
=-{\vk\/\x^2}+\Big({\e_--4s\/2\x}\Big)_t'=-{\vk\/\x^2}-\k_t'.
$$
 $\BBox$

\medskip

We begin to prove the main theorems.

\no {\bf Proof of Theorem \ref{evasEuB}.}
 Our proof is based on
the eigenvalue asymptotics \er{4g.asDir} for the operator $H$ (see
below in this section) and the Barcilon-Gottlieb transformation (see
Section 5). Let $\a,\b\in\mH_3$ and let $p,q$ have the form
\er{peb}, \er{qeb}. Then $(p,q)\in\mH_2\ts \mH_0$. Now we assume
that Theorem \ref{4g.thasdir} is valid. Then asymptotics
\er{4g.asDir} and Lemma \ref{lmEuB} yield
\[
\lb{asln1} \l_n=(\pi n)^4-2(\pi n)^2 \wh p_0+\p_1-\wh
V_{cn}+o(n^{-1}) \qqq as \qq n\to\iy,
\]
 where
\[
\lb{defQ1}
\p_1=\wh V_0-{1\/2}\int_0^1\big(p^2(t)-\wh p_0^2\big)dt,
\]
$\wh p_0,\wh q_0,\wh V_{cn}$ are given by \er{Fc},
$V=q-{p''\/2}$, and $ y_x'={dy\/dx},y'={dy\/dt}. $ Identity
\er{idp1} gives
\[
\lb{p0p0}
\wh p_0=\int_0^1p(t)dt=
\k(0)-\k(1)-\int_0^1{\vk\/\x^2}dt=-\p_0,
\]
where $\p_0$ is defined by \er{psi}.
Relations  \er{peb}, \er{qeb}
give
$$
V=q-{p''\/2}=2(\s')^2+4\s'\s^2+\s^4
-2(\vp\s)'-2\vp\s^2-
{\vp''\/2}+\s\s''+\upsilon.
$$
Let $n\to\iy$.
Using $\s''',\vp''\in L^1(0,1)$
we have
\[
\lb{fcV1}
\wh V_{cn}=\int_0^1\Big(\upsilon(t)-{\vp''(t)\/2}\Big)
\cos 2\pi ntdt+o(n^{-1}).
\]
Due to \er{vp'1} and  $\e_\pm''',\x'''\in L^1(0,1)$ we obtain
\[
\lb{asVg}
%$$
\begin{aligned}
\wh V_{cn}=\int_0^1\Big(\upsilon(t)-{\e_-'''(t)\/4\x(x(t))}\Big)
\cos 2\pi ntdt+o(n^{-1})
\\
=\int_0^1\Big( Q(x)-{(\e_-)_{xxx}'''\/4\x^4(x)}\Big)
\x(x)\cos 2\pi nt(x)dx+o(n^{-1})
=\g_n+o(n^{-1}),
\end{aligned}
\]
where $\g_n$ is defined by \er{ebg}. The substitution of \er{p0p0},
\er{asVg} into \er{asln1} yields  \er{EBasDir}. \BBox
\medskip

{\no \bf Remark}. Identity \er{defQ1} shows
that the constant $\p_1$ in asymptotics
\er{EBasDir} has the form
\[
\lb{psi1}
\p_1=\int_0^1\p(t)dt,\qqq\where\qqq
\p=V-{p^2-\wh p_0^2\/2},\qqq  V=q-{p''\/2},
\]
where $p,q$ are given by \er{peb}, \er{qeb}.
In Proposition \ref{propmAmB} we express the constant $\p_1$
in terms of $\a,\b$.

\medskip

We prove the Ambarzumyan type theorem
about the inverse problem for the operator
$\cE$.

\medskip

\no {\bf Proof of Theorem \ref{cor1}.}
Identities $a=b=1$ give $\a=\b=0$, which yields $\l_n=(\pi n)^4$.
Conversely, let $\k(0)=\k(1)$ and let
$\l_n=(\pi n)^4$ for all $n\in\N$. Asymptotics \er{EBasDir}
implies $\p_0=0$, then
\er{psi} yields $\vk=0$. Identity \er{psi}
gives $\a=\b=0$, i.e. $a=b=1$,
which proves the statement.
$\BBox$

\subsection{Eigenvalue asymptotics for the operator $H$}
Consider the operator $H$. We rewrite the equation
\[
\lb{4g.1b}
y^{(4)}+2(py')'+qy=\l y,\qqq\l\in\C,
\]
in the vector form
\[
\lb{4g.fe}
{\bf y}'-\ma
0&1&0&0\\
0&0&1&0\\
0&0&0&1\\
\l&0&0&0\\
\am {\bf y}=-\ma
0&0&0&0\\
0&0&0&0\\
0&2p&0&0\\
q&0&0&0\\
\am{\bf y},\qqq\where\qq
{\bf y}=\ma y\\ y'\\y''\\y'''+2py'\am.
\]
The corresponding matrix equation has the unique
$4\ts 4$~-~matrix valued solution
$$
M(t,\l)=\big(M_{jk}(t,\l)\big)_{j,k=1}^4,\qqq  M(0,\l)=\1_4,
$$
where $\1_4$ is the $4\ts 4$ identity matrix. The matrix valued
function $M(t,\l)$ is called the {\it fundamental matrix}. Each
function $M(t,\l),t\in\R,$ is entire in $\l$ and real at real
$\l,p,q$. The function
$M(1,\l)$ is called the {\it monodromy matrix}.

We define the determinant $D$ by
\[
\lb{defD}
D=\det\cM,\qqq
\cM(\l)=\ma M_{12}&M_{14}\\M_{32}&M_{34}\am(1,\l).
\]
The function $D$ is  entire in $\l$, since $M$ is entire.

In the unperturbed case $p=q=0$ we have $M=M_0$, where
\[
\lb{M00}
M_0(x,\l)=(M_{0,jk}(x,\l))_{j,k=1}^4
=\ma \vp_1&\vp_2&\vp_3&\vp_4\\
\l\vp_4&\vp_1&\vp_2&\vp_3\\
\l\vp_3&\l\vp_4&\vp_1&\vp_2\\
\l\vp_2&\l\vp_3&\l\vp_4&\vp_1\am(x,\l),
\]
\[
\lb{vp0} \vp_1=\vp_2',\qqq \vp_2={s_+\/z},\qqq \vp_3=\vp_4',\qqq
\vp_4={s_-\/z^3},\qqq s_\pm={\sinh zx\pm\sin zx\/2}.
\]
Here we have used the new spectral variable $z\in\C$ defined by
$$
z=\l^{1\/4},\qqq \arg z\in
S=\Big(-{\pi\/4},{\pi\/4}\Big]\qqq\as\qq\arg\l\in(-\pi,\pi].
$$
Then the unperturbed  determinant (i.e., at $p=q=0$) has the form
\[
\lb{detcM0sm}
D_0(\l)
=\det\ma\vp_2&\vp_4\\\l\vp_4&\vp_2\am(1,\l)
={\sinh z\sin z\/z^2}.
\]

The following result is well known. We give the proof for  the sake
of completeness.

\begin{lemma}
\lb{LmEvDet}
Let $p,q\in L^1(0,1)$.
Then the spectrum of $H$ satisfies the identity
\[
\lb{spch}
\s(H)=\{\l\in\C:D(\l)=0\}.
\]
Moreover, the algebraic multiplicity of each eigenvalue $\l$ of $H$
is equal to the multiplicity of $\l$  as a zero of $D$.
\end{lemma}

\no {\bf Proof.}
Any vector solution ${\bf y}$ of equation \er{4g.fe}
satisfies the identity
\[
\lb{My}
{\bf y}(1,\l)=M(1,\l){\bf y}(0,\l)\qqq\forall\qq\l\in\C.
\]
Let $\l$ be the eigenvalue of $H$
and let $y(t,\l)$ be the corresponding eigenfunction.
Then identities \er{4g.dc} give
$$
{\bf y}(0,\l)=(0, g_0,0,h_0)^\top,\qqq
{\bf y}(1,\l)=(0, g_1,0,h_1)^\top
$$
where $g_j=y'(j,\l)$ and $h_j=y'''(j,\l)+2p(j)y'(j,\l)$ for $j=0,1$.
Identity \er{My} implies
\[
\lb{sysD} M(1,\l)(0, g_0,0,h_0)^\top =(0, g_1,0,h_1)^\top.
\]
This  formula yields $D(\l)=0$. Conversely, if $D(\l)=0$, then the
system \er{sysD} has the non-trivial solution $(g_0,h_0)$, which
gives the eigenfunction $y(t,\l)$. This proves \er{defD}.

Let $\l$ be an eigenvalue of $H$.
It is well known (see, e.g., \cite[Ch.~I.2.3.VI]{Na})
that the total dimension of the root subspace of $\l$,
i.e.  the algebraic multiplicity of $\l$,
coincides with the multiplicity of $\l$ as a zero of $D$.
$\BBox$

\medskip

Recall that the spaces $\mH_m$ are given by \er{mHm}. Below  we
assume, without loss of generality, that $q$ belongs to the Sobolev
space $\mH_0^0$, where $\mH_m^{0}, m\ge 0$ are given  by
$$
 \mH_m^{0}=\Big\{f\in \mH_m:\int_0^1f(t)dt=0\Big\}.
$$

\begin{lemma}
\lb{LmrAsF} Let $(p,q)\in \mH_2\ts \mH_0^0$. Then the function $D$,
defined  by \er{defD}, satisfies
\[
\lb{asF5}
D(\l)
=D_0(\l)+e^{\Re z+|\Im z|}{O(1)\/z^3},
\]
as $|\l|\to\iy,z\in S$, uniformly on bounded subsets
of $\mH_2\ts \mH_0^0$, where $D_0$ is given by \er{detcM0sm}.
\end{lemma}

\no  The proof of this Lemma see in Section 7. It is based on
Birkhoff's approach, see, e.g.,\cite{Na}.

\medskip

We prove the Counting Lemma for the determinant $D(\l)$.

\begin{lemma}
\lb{lmCl} Let $(p,q)\in \mH_2\ts \mH_0^0$. For any integer $N\ge 1$
large enough the entire function $D(\l)$ has exactly $N$ zeros,
counting with multiplicities, in the disc
$\{|\l|<\pi^4(N+{1\/2})^4\}$ and for each integer $n>N$ it has
exactly one simple zero in the domain $\{|z-\pi n|<{\pi\/4}\}$.
There are no other zeros.
\end{lemma}

\no {\bf Proof.} Let $N\in\N$ be large enough and let $N'>N$ be an
integer. Let $\l\in\C$ belong to the contours
\[
\lb{contours} \textstyle |z|=\pi\Big(N+{1\/2}\Big),\qqq
|z|=\pi\Big(N'+{1\/2}\Big),\qqq |z-\pi n|={\pi\/4},\qq n>N.
\]
Then the estimates $|\sinh z|>{e^{\Re z}\/4}$ and $|\sin z|>
{e^{|\Im z|}\/4}$ and asymptotics \er{asF5} give
$$
\Big|D(\l)-D_0(\l)\Big|<|D_0(\l)|
$$
on all contours.
Hence, by Rouch\'e's theorem, the function $D$
has the same number of zeros as the function $D_0$
in each bounded domains and in the remaining
unbounded domain. Since $D_0$ has exactly one simple
zero at each point
$\l=(\pi n)^4,n\in\N$, and since
$N'>N$ can be chosen arbitrary large,
the Lemma follows.
$\BBox$

\medskip
We need the  sharper asymptotics of the function $D$.

\begin{lemma}
\lb{Lmascf}
Let $(p,q)\in \mH_2\ts \mH_0^0$.

i) Let $|\l|\to\iy,z\in S$. Then the function $D(\l)$ satisfies
\[
\lb{4g.adwM12}
D(\l)=E(z)\Big(\sin w(z)
+{p'(1)-p'(0)\/8z^3}\cos w(z)
+{F(z)\/z^3}
\Big),
\]
where the function $F(z)$ is analytic in $S$ and
\[
\lb{4g.defwj}
E(z)={e^{z-{\wh p_0\/2z}}\/2z^2},\qqq
w(z)=z+{\wh p_0\/2z},
\]
\[
\lb{estF} |F(z)|\le Ce^{|\Im z|}\qqq\forall\qq z\in \{S:|z|>r\},
\]
for some  $C>0$ and $r>0$ large enough.

ii) Let $z_n=\l_n^{1\/4}\in S,n\in\N$, where $\l_n$ are
eigenvalues of $H$. Then
\[
\lb{asznr}
w(z_n)=\pi n+O(n^{-3})\qqq\as\qq n\to\iy
\]
uniformly on bounded subsets of $\mH_2\ts \mH_0^0$.

iii) Let $(p,q)\in \mH_{2+j}\ts \mH_j^0,j\in\{0,1\}$ and let
$n\to\iy$. Then
\[
\lb{aswF1}
F(z_n)={(-1)^n\/8}\Big(\int_0^1p^2(t)dt+2\wh V_{cn}\Big)+\wt F_n,
\]
where
\[
\lb{awF1} \wt F_n=\ca o(n^{-1})  & if\ j=0\\ O(n^{-2}) &  if\
j=1\ac
\]
uniformly on bounded subsets of $\mH_{2+j}\ts \mH_j^0$.

\end{lemma}

\no{\bf Proof.} Proof of i) and iii) is given in Section 7.

ii) Asymptotics \er{4g.adwM12} implies
\[
\lb{aswMr}
D(\l)=E(z)\Big(\sin w(z)+{O(e^{|\Im z|})\/|z|^3}\Big)\qqq
\as\qq|z|\to\iy,\qq z\in S.
\]
Lemma \ref{lmCl} gives $|z_n-\pi n|< {\pi\/4}$ for all $n\in\N$
large enough. If $n\to\iy$, then the second identity in
\er{4g.defwj} implies
\[
\lb{zd1} w(z_n)=\big(1+{\wh p_0\/2z_n^2}\big)z_n=\pi n+\ve_n,\qq
|\ve_n|<1.
\]
The substitution this asymptotics into \er{aswMr} and the identity
\er{spch}  $D(\l_n)=0$ give
\[
\lb{asF1} D(\l_n)=(-1)^nE(z_n) \big(\sin\ve_n+O(n^{-3})\big),
\]
which yields $\ve_n=O(n^{-3})$, since $|\ve_n|<1$. Thus
asymptotics \er{zd1} implies \er{asznr}. $\BBox$

\medskip

We determine high energy eigenvalue asymptotics for the operator
$H$.
\medskip

\no {\bf Proof of Theorem \ref{4g.thasdir}.} Let $n\to\iy$. The
asymptotics  $w(z_n)=\pi n+\ve_n$ and $\ve_n=O(n^{-3})$ from
\er{asznr} and \er{4g.adwM12} imply
\[
\lb{asF11} D(\l_n) =(-1)^nE(z_n)\Big( \ve_n+{p'(1)-p'(0)\/8(\pi
n)^3} +{(-1)^nF(z_n)\/(\pi n)^3}+{O(1)\/n^7}\Big).
\]
Asymptotics \er{aswF1}, \er{awF1}, identities \er{asF11} and
$D(\l_n)=0$ yield
\[
\lb{zd2} \ve_n=-{P\/8(\pi n)^3} -{\wh V_{cn}\/4(\pi
n)^3}+\wt\ve_n,\qqq \wt\ve_n=\ca o(n^{-4}), j=0\\ O(n^{-5}), j=1\ac,
\]
where $P=\int_0^1(p^2+p'')dt$. Identities \er{4g.defwj} and
asymptotics  \er{asznr} give
$$
z_n+{\wh p_0\/2z_n}=\pi n-{P\/8(\pi n)^3} -{\wh V_{cn}\/4(\pi
n)^3}+\wt\ve_n,
$$
and then
\[
\lb{zd3} z_n=\pi n-{\wh p_0\/2\pi n} -{P+2\wh p_0^2\/8(\pi
n)^3}-{\wh V_{cn}\/4(\pi n)^3} +\wt\ve_n+{O(1)\/n^5},
\]
which yields \er{4g.asDir}. $\BBox$

\section{Examples and remarks }
\setcounter{equation}{0}

\subsection{Fourth order operator}
We consider few examples of the operator $H$.

\no {\bf Example 1.} Here we determine the eigenvalue asymptotics
for the operators with $\d$-coefficients. Eigenvalue asymptotics for
the even order operators $\pa^{2n}+q$ on the circle, where $q$ is a
distribution, were considered by Mikhailets, Moliboga \cite{MM}.

We consider the boundary value problem \er{4g.dc} for the operator
$H=\pa^4+\g\d(t-t_0)$, where $t_0\in(0,1),\g\in\C$, $\d(t)$ is the
standard Dirac $\d$-function. In this case $p=0,q=\g\d(t-t_0)$.
Note that our forth order operators with the $\d$-coefficients
can be interpreted in the sense of forms as well as the
second order operators, see \cite[Ch X.2]{RS}.
The
usual formulas for the fundamental solutions (see, e.g., \cite{BK1})
give
$$
\cM(\l)=\ma\vp_2&\vp_4\\\l\vp_4&\vp_2\am(1,\l)
-{\g\/z^2}\ma z^{-2}s_-(1-t_0)s_+(t_0)&
-z^{-4}s_-(1-t_0)s_-(t_0)\\
s_+(1-t_0)s_+(t_0)&
z^{-2}s_+(1-t_0)s_-(t_0)
\am,
$$
where
$s_\pm$ are given by \er{vp0}.
Then the determinant $D$ has the form
\[
\lb{deltadetcM}
D=D_0
+{\g\/4z^5}\Big(\big(\cos(2t_0-1)z-\cos z\big)\sinh z
+\big(\cosh(2t_0-1)z-\cosh z\big)\sin z\Big).
\]
Identity $D(\l_n)=0$ provides the eigenvalue asymptotics
$$
\l_n=(\pi n)^4+\g(1-\cos2\pi nt_0)+O(n^{-3})\qqq\as\qq n\to\iy
$$
uniformly in $\g$ on bounded subsets of $\C$. If $t_0={1\/2}$, then
the coefficient $q=\g\d(t-{1\/2})$ is even and \er{deltadetcM} gives
$$
\l_{2n}=(2\pi n)^4\qq\forall\qq n\in\N,\qqq
\l_{2n-1}=\big((2n-1)\pi\big)^4+2\g+O(n^{-3})\qq\as\qq n\to\iy
$$
uniformly on bounded subsets of $\C$.

\medskip

\no {\bf Example 2.} Consider the boundary value problem \er{4g.dc}
for the even case
$$
\textstyle H=\pa^4+\g\pa\d(t-{1\/2})\pa,\qq \g\in\C, \qqq i.e.,\qq
p=\g\d(t-{1\/2}),\qq q=0.
$$
The determinant has the form
$$
D=D_0
+{\g\/2z^3}(\sinh z\cos z-\cosh z\sin z)
+{\g^2\/8z^4}(1-\cosh z\cos z).
$$
This identity implies the following asymptotics
$$
D(\l)={e^z\/2z^2}\Big(\sin z+
{\g\/2z}(\cos z-\sin z)-{\g^2\/8z^2}\cos z
+O(e^{-\Re z})\Big)\qq\as\qq |z|\to\iy,\qq z\in S,
$$
uniformly in $\g$ on bounded subsets of $\C$.
Identity $D(\l_n)=0$ gives the eigenvalue asymptotics
$$
\l_n=(\pi n)^4-2\g (\pi n)^2-{\g^2\/2}\pi n+{\g^2\/2}-{\g^3\/12}+{O(1)\/n^2}
\qqq\as\qq n\to\iy
$$
uniformly on bounded subsets of $\C$.

\medskip

\no {\bf Example 3: the square of a second order operator.} Consider
the operator
$$hy=-y''-py,\qqq y(0)=y(1)=0, \qq \qqq p\in\mH_4,
$$
on the unit interval $ [0,1]$.
 Let $\a_n,n\in\N$, be eigenvalues of this operator labeled by $
|\a_{1}|\le|\a_{2}|\le|\a_{3}|\le... $ counted with multiplicity. It
is well known (see \cite[(4.21)]{FP}), that
\[
\lb{asaln}
\a_n=(\pi n)^2-\wh p_0+
{1\/(2\pi n)^2}\Big(\int_0^1(p^2+p'')dt-\wh p_0^2\Big)
+{O(1)\/n^4}\qqq\as\qq n\to\iy.
\]
This gives the following asymptotics for the eigenvalues
$\l_n=\a_n^2$ of the operator $h^2=\pa^4+2\pa p\pa +q$, where
$q=p''+p^2$:
\[
\lb{asaln2}
\l_n=\a_n^2
=(\pi n)^4-2(\pi n)^2\wh p_0+{1\/2}\Big(\int_0^1(p^2+p'')dt +\wh
p_0^2\Big)+{O(1)\/n^2}
\]
as $n\to\iy$. Asymptotics \er{asaln2} is in agreement with
\er{4g.asDir}, since $V=q-{p''\/2}=p^2+{p''\/2}$ in our case and
then
$$
\wh V_0-{1\/2}\int_0^1(p^2-\wh p_0^2)dt
={1\/2}\Big(\int_0^1(p^2+p'')dt
+\wh p_0^2\Big).
$$

\medskip

\no {\bf Remark 1.} Fulton and Pruess \cite[(4.21)]{FP} determined
the correct asymptotics
\[
\lb{asFP}
\a_n=(\pi n)^2-\wh p_0+
{P-\wh p_0^2\/(2\pi n)^2}
+{Q_2\/(2\pi n)^4} +{O(1)\/n^6},
\]
where $P=\int_0^1(p^2+p'')dt$,
$$
Q_2=2\wh p_0(12P-11\wh p_0^2)-2\int_0^1p^3(s)ds+\int_0^1{p'}^2(s)ds
-\big(6pp'+p'''\big)\big|_0^{1}.
$$

\no {\bf Remark 2.} P\"oshel and Trubowitz,
see \cite[Problem~2.3]{PT},
determined asymptotics
$$
\a_n=(\pi n)^2+
{1\/(2\pi n)^2}\int_0^1(p^2+p'')dt+{o(1)\/n^2},
$$
in the class
of real $p,p''\in L^2(0,1),\wh p_0=0$ (note that  there is a misprint
in the sign in the second term of asymptotics in \cite{PT}).

\medskip

\no {\bf Remark 3.} Dikii \cite[p.~189]{D1} considered an operator
$h$ with the real $p\in C^\iy[0,1]$ and determined the following
asymptotics
\[
\lb{asq_1D}  \a_n=(\pi n)^2-\wh p_0 +{1\/(2\pi n)^2}
\Big(\|p\|^2-4\wh p_0^2+{1\/3}\big(p'(1)-p'(0)\big)\Big)+...,
\]
This also is in a disagreement with \er{asaln}.
The coefficients
$4$ and ${1\/3}$ in \er{asq_1D} are mistaken,
see \cite[Remark 4.5]{FP}.

\medskip

\no {\bf Remark 4.} Sadovnichii \cite[p.~308-309]{S} considered the
operator $H=\pa^4+2\pa p\pa +q$, where
\[
\lb{Sc}
p,q\in C_\R^\iy[0,1],\qqq
p^{(j)}(0)=p^{(j)}(1)=q^{(2j-1)}(0)=q^{(2j-1)}(1)=0\qqq
\forall\qq j\in\N,
\]
Let $\l_n,n\in\N$, be eigenvalues of this operator
labeled by $\l_1\le\l_2\le ...$ counted with multiplicities.
Sadovnichii wrote (without the proof) the following asymptotics
\[
\lb{asS} \l_n=(\pi n)^4-2(\pi n)^2\wh p_0+\wh
q_0+{c_1\/n^2}+{c_2\/n^4}+...
\]
as $n\to\iy$,
where $c_1,c_2$ are some undetermined constants.
Consider the operator
$h^2=(-\pa^2-p)^2=\pa^4+2\pa p\pa +p''+p^2$, where $p$  satisfies \er{Sc}.
In this case asymptotics  \er{asS}
gives
\[
\lb{asS1} \l_n=(\pi n)^4-2(\pi n)^2\wh
p_0+\|p\|^2+{c_1\/n^2}+{c_2\/n^4}+...
\]
On the other hand, asymptotics \er{asaln2} of $\a_n^2$ yields in this
case
\[
\lb{asFP1} \l_n=\a_n^2=(\pi n)^4-2(\pi n)^2\wh p_0+{\|p\|^2+\wh
p_0^2\/2}+{O(1)\/n^2}.
\]
The third term in asymptotics
\er{asS1} is in a disagreement with the corresponding term in
\er{asFP1}.
Therefore, the term $\wh q_0$ in \er{asS} is incorrect.

\subsection{The Euler-Bernoulli operator}
We give some remarks and examples for the Euler-Bernoulli operator.

\no {\bf Remark 5.} Jian-jun, Kui and  Da-jun \cite{JKD}
considered the
Euler-Bernoulli equation $(au'')''=\l b u$, where $a,b$
are smooth positive coefficients, under the following four boundary
conditions:
\[
\lb{cfb}
u(0)=u'(0)=u''(1)=(au'')'|_{x=1}=0\qqq\text{(clamped-free beam)},
\]
\[
\lb{csb}
u(0)=u'(0)=u'(1)=(au'')'|_{x=1}=0\qqq\text{(clamped-sliding beam)},
\]
\[
\lb{cpb}
u(0)=u'(0)=u(1)=u''(1)=0\qqq\text{(clamped-pinned beam)},
\]
\[
\lb{ccb}
u(0)=u'(0)=u(1)=u'(1)=0\qqq\text{(clamped-clamped beam)}.
\]
In fact, they considered the four operators. For these four
operators they announced the following asymptotics
\[
\lb{asChin}
\l_n=a_n^4+4a_n^2
\int_0^1\Big({5\x''\/4\x^2}-{15(\x')^2\/8\x^3} -{3(a')^2\/8\x
a^2}+{a''\/2a\x}\Big)dx +O(1)
\]
as $n\to\iy$,
where
\[
\lb{defan}
a_n=\ca \pi(n-{1\/2})\ \text{for conditions}\ \er{cfb}\\
\pi(n-{1\/4})\ \text{for conditions}\ \er{csb}\\
\pi(n+{1\/4})\ \text{for conditions}\ \er{cpb}\\
\pi(n+{1\/2})\ \text{for conditions}\ \er{ccb}
\ac,
\]
 see \cite[(3.8)]{JKD}.
Direct calculations show that
$$
2\int_0^1\Big({5\x''\/4\x^2}-{15(\x')^2\/8\x^3}
-{3(a')^2\/8\x a^2}+{a''\/2a\x}\Big)dx=\p_0,
$$
where $\p_0$ is given by \er{psi}.
Then, assuming that asymptotics \er{asChin} are correct,
we obtain
\[
\lb{evaschin}
\l_n=a_n^4+2a_n^2\p_0+O(1)
\]
as $n\to\iy$. This asymptotics
allows to extend the results of Theorem \ref{cor1} onto the
Euler-Bernoulli operator with the boundary conditions
\er{cfb}--\er{ccb}.

\begin{proposition} Let $Q=0$ and let
the real $(\a,\b)\in \mH_3\ts\mH_3$ satisfy the
conditions \er{sigmv} and $\k(0)=\k(1)$.
Assume that eigenvalues $\l_n,n\in\N$, of the Euler-Bernoulli
operator \er{defcE} under one of the boundary conditions
\er{cfb}--\er{ccb}
satisfy
asymptotics \er{asChin} with the coefficients $a_n$
given by \er{defan}.
 Then the eigenvalues
$\l_n=(\pi n)^4$ for all $n\ge 1$ iff $a=b=1$.
\end{proposition}

\no {\bf Proof.}  Using \er{evaschin} and
repeating the arguments from the proof
of Theorem \ref{cor1} we obtain the statement.
\BBox

\medskip

\no {\bf Remark 6.} Consider the weighted second order operator
$h_w$ given by
\[
\lb{soo} h_wu=-{1\/b}u_{xx}'',\qqq x\in[0,1],\qqq u(0)=u(1)=0,
\]
with real periodic coefficients $b\in\mH_2, b>0$, normalized by
$\int_0^1b^{1\/2}ds=1$. Let $\l_1<\l_2<...$ be eigenvalues of this
operator. The standard Liouville transformation
$$
t(x)=\int_0^xb(s)^{1\/2}ds,\qqq
y(t)=b^{1\/4}(x(t))u(x(t)),
$$
yields that operator $h_w$ is unitarily equivalent to the operator
$\wt h$ given by
$$
\wt hy=-y_{tt}''+(\b^2+\b_t')y,\qqq t\in[0,1],\qqq y(0)=y(1)=0,
$$
where $\b={b'\/4b}$. The standard eigenvalue asymptotics for the
second order operator gives
$$
\l_n=(\pi n)^2+\int_0^1\b^2(t)dt+o(1)\qqq\as\qq n\to\iy,
$$
 see \cite{PT}.
Assume that $\l_n=(\pi n)^2$ for all $n\in\N$. Then the last
asymptotics implies $\b=0$, which yields  $b=1$. We obtain a result,
similar to the result of Theorem \ref{cor1} for the operator
\er{soo}: the eigenvalues $\l_n=(\pi n)^2$ for all $n\in\N$ iff
$b=1$.

\medskip

We need the following identity for the constant $\p_1$
given by \er{psi1}, see the proof in Section 5.

\begin{proposition}
\lb{propmAmB}
Let $\a,\b\in\mH_3$.
Then
\[
\lb{defQ}
\p_1=\mA(1)-\mA(0)+\int_0^1\mB(x)\x(x)dx +{\p_0^2\/2}+Q_0,
\]
where $\p_0$ is given by \er{psi},
$Q_0=\int_0^1Q(x)\x(x)dx$
and the functions $\mA(x),\mB(x)$ have the form
\[
\lb{A2}
\mA={1\/\x^3}\Big({2s^3\/3}-{\e_-^3\/2}-2\e\e_+
-(s-\e_-)\e_- s
+(s-\e_-)_x's-(\a\e_-)_x'
-{(\e_-)_{xx}''\/4}\Big),
\]
\[
\lb{B2}
\mB={1\/8\x^4}\Big(\big((\e_-)_x'-\e_-^2-2\vk\big)^2
-8\big((\e_+)_x'-2\e\big)^2\Big),
\]
$\vk$ and $s,\e_\pm,\e$ are given by \er{psi} and \er{sphi},
respectively.
\end{proposition}

{\bf Example 4: the Euler-Bernoulli operators
with periodic coefficients.}
Consider the case of 1-periodic $a,b$.
Let $\a,\b,\a''',\b'''\in L^1(\R/\Z)$ satisfy \er{sigmv}
and let $Q=0$.
Identities \er{psi}, \er{defQ} give
\[
\lb{Qpper}
\p_0=\int_0^1{\vk(x)\/\x(x)}dx>0,\qqq
\p_1=\int_0^1\mB(x)\x(x)dx+{\p_0^2\/2}.
\]
Consider some particular cases.

{\bf Example 4.1.} Let, in addition, $ab=1$.
In this case we have $\a=-\b$ and
identities \er{sigmv}, \er{psi}, \er{sphi}, \er{B2} give
$$
\x=a^{-{1\/2}}>0,\qq \e_-=-2\a,\qq\e_+=0,\qq
s=-\a,\qq\vk=\a^2,
\qq
\mB={1\/2}a^2(\a'+6\a^2)^2.
$$
The Euler-Bernoulli operator
$$
\cE u=a(au_{xx}'')_{xx}''=h_a^2u,
$$
where the operator $h_au=-au_{xx}''$ with the boundary condition
$u(0)=u(1)=0$. Identities \er{Qpper} yield
$$
\p_0=\int_0^1a^{1\/2}\a^2dx,\qqq
\p_1={1\/2}\int_0^1a^{3\/2}\big(\a'+6\a^2\big)^2dx
+{1\/2}\Big(\int_0^1a^{1\/2}\a^2dx\Big)^2\ge 0.
$$
Identity \er{ebg} gives
$$
\g_n={1\/2}\int_0^1 a^{3\/2}(x)\a'''(x)
\cos\Big(2\pi n\int_0^x {ds\/a^{1\/2}(s)}\Big)dx.
$$
Substituting these identities into \er{EBasDir}
we obtain the eigenvalue asymptotics for this case.
Note that the eigenvalues $\sqrt{\l_n}$ of the operator
$-au_{xx}'',u(0)=u(1)=0$, satisfy
$$
\sqrt{\l_n}=(\pi n)^2+\p_0+{\p_1-\p_0^2-\g_n\/2(\pi n)^2}
+{o(1)\/n^3}\qqq\as\qq n\to\iy.
$$

{\bf Example 4.2.} Consider the case $a=1$.
Then $\a=0$ and identities
\er{sigmv}, \er{psi}, \er{sphi}, \er{B2} give
$$
\x=b^{1\/4}>0,\qqq\e_+=
\e_-=\b,\qqq
\vk={5\/4}\b^2,\qqq
\mB=-{1\/8b}\Big(7(\b')^2-25\b'\b^2+{79\/4}\b^4\Big).
$$
Identities \er{Qpper} yield
$$
\p_0={5\/4}\int_0^1{\b^{2}\/b^{1\/4}}dx,\qqq
\p_1=-{1\/8}\int_0^1{1\/b^{3\/4}}
\Big(7(\b')^2-25\b'\b^2+{79\/4}\b^4\Big)dx
+{25\/32}\Big(\int_0^1{\b^{2}\/b^{1\/4}}dx\Big)^2.
$$
Identity \er{ebg} gives
$$
\g_n=-{1\/4}\int_0^1 \b'''(x)
\cos\Big(2\pi n\int_0^x b^{1\/4}(s)ds\Big){dx\/b^{3\/4}(x)}.
$$
Substituting these identities into \er{EBasDir}
we obtain the eigenvalue asymptotics for this case.

{\bf Example 4.3.} Consider the operator
$$
\cE u={1\/a}(au_{xx}'')_{xx}''.
$$
Then $a=b$ and $\a=\b$. Identities
\er{sigmv}, \er{psi}, \er{sphi}, \er{B2}
give
$$
\x=1,\qqq \e_+=2\a,\qqq \e_-=0,\qqq\vk=4\a^2,
\qqq
\mB=-4\big((\a')^2-2\a^4\big).
$$
Identities \er{Qpper} yield $\p_0=4\int_0^1\a^2dx=4\|\a\|^2,$
$$
\p_1=-4\int_0^1\big((\a')^2-2\a^4\big)dx
+8\Big(\int_0^1\a^2dx\Big)^2=8(\|\a^2\|^2+\|\a\|^4)-4\|\a'\|^2,
$$
where $\|f\|^2=\int_0^1f^2dx$. Depending on $\a$, the constant
$\p_1$ may be positive or negative. The definition \er{ebg} gives
$$
\g_n=0.
$$
Substituting these identities into \er{EBasDir}
we obtain
$$
\l_n=(\pi n)^4+8\|\a\|^2(\pi n)^2+8(\|\a^2\|^2+\|\a\|^4)-4\|\a'\|^2
+{o(1)\/n}\qqq\as\qq n\to\iy.
$$

\section{Euler-Bernoulli operators with near constant coefficients}
\setcounter{equation}{0}

\subsection{Fundamental matrix.}
We consider the operator $\cE_\ve$, given by \er{cEsm}, for small
$\ve$. Rewrite the equation
\[
\lb{eqsm}
c_\ve (a^\ve u'')''+\ve Q u=\l b^\ve u
\]
in the vector form
\[
\lb{4g.fesm}
{\bf u}'=\cC{\bf u},
\]
where
\[
\lb{ucC}
{\bf u}=\ma u\\u'\\c_\ve a^\ve u''\\c_\ve(a^\ve u'')'\am,\qqq
\cC=\ma 0&1&0&0\\0&0&(c_\ve a^{\ve})^{-1}&0\\0&0&0&1\\
\l b^\ve-\ve Q&0&0&0\am.
\]
The {\it fundamental matrix}
$M(x,\l,\ve)=(M_{jk}(x,\l,\ve))_{j,k=1}^4,
(x,\l,\ve)\in[0,1]\ts\C\ts\R$,
of equation \er{eqsm}
satisfies
\[
\lb{mesm}
M'=\cC M,\qqq M(0,\l,\ve)=\1_4.
\]

In the unperturbed case $\ve=0$ we have $M(x,\l,0)=M_0(x,\l)$,
where $M_0$ is given by \er{M00}.

\begin{lemma}
\lb{LmMMsm}
Let $\a,\b,Q\in L^1(0,1)$.
 Then each matrix valued function $M(x,\l,\cdot), (x,\l)\in[0,1]\ts\C$
is analytic in $\ve\in \{\ve\in \C:|\ve|<\ve_1\}$ for some
$\ve_1>0$, and satisfies
\[
\lb{asMsm}
M(x,\l,\ve)=M_0(x,\l)+\ve M_1(x,\l)+O(\ve^2)
\]
as $\ve\to 0$
uniformly in $(x,\l)$ on bounded subsets of $[0,1]\ts\C$,
where the matrix $M_1=(M_{1,jk})_{j,k=1}^4$ has the form
\[
\lb{M1jk}
M_{1,jk}(x,\l)=\int_0^x\Big(\a_1(t)M_{0,j2}(x-t,\l)M_{0,3k}(t,\l)
+\b_1(t,\l)M_{0,j4}(x-t,\l)M_{0,1k}(t,\l)\Big)dt,
\]
\[
\lb{a1b1sm}
\a_1=-\ln b(0)-4\int_0^x\a(s)ds
-4\int_0^1\int_0^x\big(\b(s)-\a(s)\big)dsdx,
\qq
\b_1=4\l \int_0^x\b(s)ds+\l\ln b(0)- Q.
\]

\end{lemma}

\no {\bf Proof.} Identities \er{abxe}, \er{cvesm}  give
$$
\begin{aligned}
a^\ve(x)=1+4\ve \int_0^x\a(s)ds+O(\ve^2),\qqq
b^\ve(x)=1+\ve\ln b(0)+4\ve\int_0^x\b(s)ds+O(\ve^2),
\\
c_\ve=1+\ve\ln b(0)+4\ve\int_0^1\int_0^x\big(\b(s)-\a(s)\big)dsdx+O(\ve^2)
\end{aligned}
$$
as  $\ve\to 0$ uniformly on $[0,1]$. Substituting these asymptotics
into \er{ucC} we obtain
\[
\lb{asCsm}
\cC(x,\l,\ve)=\cC_0(\l)+\ve\wt\cC(x,\l,\ve),\qqq
\wt\cC(x,\l,\ve)=\cC_1(x,\l)+O(\ve),
\]
as $\ve\to 0$
uniformly in $(x,\l)$ on bounded subsets of $[0,1]\ts\C$,
where
\[
\lb{C01sm}
\cC_0=\ma 0&1&0&0\\0&0&1&0\\0&0&0&1\\\l&0&0&0\am,
\qqq\cC_1=(\cC_{1,jk})_{j,k=1}^4=\ma 0&0&0&0\\0&0&\a_1&0\\0&0&0&0\\
\b_1&0&0&0\am,
\]
Using asymptotics \er{asCsm} and \er{mesm} we obtain
$$
M'-\cC_0M=\ve\wt\cC M,
$$
which yields
$$
M(x,\l,\ve)=M_0(x,\l)+
\ve\int_0^xM_0(x-t,\l)\wt\cC(t,\l,\ve)M(t,\l)dt.
$$
The standard iterations show that $M$
is analytic in $|\ve|<\ve_1$ and satisfies \er{asMsm}, where
$$
M_{1,jk}(x,\l)=\int_0^x\sum_{\ell,m=1}^4
M_{0,j\ell}(x-t,\l)\cC_{1,\ell m}(t,\l)M_{0,mk}(t,\l)dt.
$$
Substituting here \er{C01sm}
we obtain \er{M1jk}.
$\BBox$

\subsection{Determinant.}
Introduce the determinant
\[
\lb{Dsm}
D(\l,\ve)=\det\cM(\l,\ve),\qqq
\cM(\l,\ve)=\ma M_{12}&M_{14}\\M_{32}&M_{34}\am(1,\l,\ve).
\]
The arguments from the proof of
Lemma \ref{LmEvDet} show that
the spectrum of $\cE_\ve$ satisfies the identity
\[
\lb{spchcE}
\s(\cE_\ve)=\{\l\in\C:D(\l,\ve)=0\}.
\]
Moreover, the algebraic multiplicity of each eigenvalue $\l$ of
$\cE_\ve$ is equal to the multiplicity of $\l$ as a zero of $D$.

\begin{lemma}
\lb{LmDetsm}
Let $\a,\b,Q\in L^1(0,1)$ and let $\ve\in\R$ be small
enough. Then

i) The determinant $D$, given by \er{Dsm}, satisfies
\[
\lb{asDsm1}
D(\l,\ve)=D_0(\l)+\ve D_1(\l)+O(\ve^2)
\]
uniformly in  bounded $\l\in\C$, where $D_0$ is given by
\er{detcM0sm} and $D_1$ is an entire function
given by
\[
\lb{D1sm}
D_1(\l)=\int_0^1\g_1(x,\l)g(x,\l)dx,
\qqq
\g_1(x,\l)=\a_1(x)+{\b_1(x)\/\l},
\]
\[
\lb{gesm}
g={1\/4z}\Big( \big(\cosh z-\cosh z(1-2x)\big)\sin z
+\big(\cos z-\cos z(1-2x)\big)\sinh z\Big).
\]
Moreover,
\[
\lb{D1pim} D_1((\pi n)^4)={(-1)^{n+1}\sinh \pi n\/2(\pi n)^2}
\Big(\wh\a_{sn}-\wh\b_{sn}
+{ \wh Q_0 -\wh Q_{cn}\/2(\pi n)^3}\Big)
\qqq\forall\qq n\in\N.
\]

ii)  The function $D(\cdot,\ve)$ has exactly one simple zero in each
domain $|\l^{1\/4}-\pi n|<1, \ n\in \N$.
\end{lemma}

\no {\bf Proof.}
i) Asymptotics \er{asMsm} implies that the matrix $\cM$,
given by \er{Dsm}, satisfies
\[
\lb{cMsm}
\cM(\l,\ve)=\cM_0(\l)+\ve \cM_1(\l)+O(\ve^2)
\]
as $\ve\to 0$ uniformly on bounded subsets of $\C$,
where, due to \er{M00},
\[
\lb{cM01}
 \cM_0=(\cM_{0,jk})_{j,k=1}^2
=\ma\vp_2&\vp_4\\\l\vp_4&\vp_2\am(1),\qq
\cM_1=(\cM_{1,jk})_{j,k=1}^2 =\ma
M_{1,12}&M_{1,14}\\M_{1,32}&M_{1,34}\am(1),
\]
and $M_1$ has the form \er{M1jk},
here and below $M(x)=M(x,\l),...$
Substituting asymptotics \er{cMsm} into definition \er{Dsm}
we obtain \er{asDsm1},
where $D_1$ is an entire function, given by
\[
\lb{D01sm}
D_1=\det\ma \cM_{0,11}&\cM_{1,12}\\\cM_{0,21}&\cM_{1,22}\am
+\det\ma \cM_{1,11}&\cM_{0,12}\\\cM_{1,21}&\cM_{0,22}\am.
\]

We prove \er{D1pim}. Identity \er{M1jk} and the definition of
$\cM_1$ in \er{cM01} yield
\[
\lb{cM1sm}
\cM_1=
\int_0^1\big(\a_1(x)\cA(x,\l)+\b_1(x,\l)\cB(x,\l)\big)dx,
\]
where $\a_1,\b_1$ are given by \er{a1b1sm},
$$
\begin{aligned}
\cA=\ma
M_{0,12}(1-x)M_{0,32}(x)&M_{0,12}(1-x)M_{0,34}(x)\\
M_{0,32}(1-x)M_{0,32}(x)&M_{0,32}(1-x)M_{0,34}(x)
\am,
\\
\cB=\ma M_{0,14}(1-x)M_{0,12}(x)&M_{0,14}(1-x)M_{0,14}(x)\\
M_{0,34}(1-x)M_{0,12}(x)&M_{0,34}(1-x)M_{0,14}(x)
\am.
\end{aligned}
$$
Using \er{M00} we obtain
\[
\lb{cAsm}
\cA=\ma
\l\vp_2(1-x)\vp_4(x)
&\vp_2(1-x)\vp_2(x)\\
\l^2\vp_4(1-x)\vp_4(x)
&\l\vp_4(1-x)\vp_2(x)
\am,
\]
\[
 \lb{cBsm}
\cB=\ma
\vp_4(1-x)\vp_2(x)
&\vp_4(1-x)\vp_4(x)\\
\vp_2(1-x)\vp_2(x)
&\vp_2(1-x)\vp_4(x)
\am={1\/\l}\ma
\cA_{22}&{1\/\l}\cA_{21}\\
\l\cA_{12}&\cA_{11}
\am.
\]
Identities \er{D01sm}, \er{cM1sm}
give
\[
\lb{D1sm1}
D_1=\int_0^1\big(\a_1(x)g(x)+\b_1(x)\wt g(x)\big)dx,
\]
where
$$
g=\det\ma \cM_{0,11}&\cA_{12}\\\cM_{0,21}&\cA_{22}\am
+\det\ma \cA_{11}&\cM_{0,12}\\\cA_{21}&\cM_{0,22}\am,
\qq
\wt g=\det\ma \cM_{0,11}&\cB_{12}\\\cM_{0,21}&\cB_{22}\am
+\det\ma \cB_{11}&\cM_{0,12}\\\cB_{21}&\cM_{0,22}\am.
$$
Using \er{cM01}, \er{cBsm} we obtain
\[
\lb{fgsm}
g=\det\ma \vp_2(1)&
\cA_{12}(x)\\
\l\vp_4(1)&\cA_{22}(x)\am
+\det\ma \cA_{11}(x)&\vp_4(1)\\
\cA_{21}(x)&\vp_2 (1)\am=\det\ma \vp_2(1)&
\l\cA_{12}(x)+\cA_{21}(x)\\
\vp_4(1)&\cA_{11}(x)+\cA_{22}(x)\am,
\]
\[
\lb{g1sm5}
\wt g={1\/\l}\Big(\det\ma \vp_2(1)&
{1\/\l}\cA_{21}(x)\\
\l\vp_4(1)&
\cA_{11}(x)\am
+\det\ma \cA_{22}(x)&\vp_4(1)\\
\l\cA_{12}(x)&\vp_2(1)\am\Big)={g\/\l}.
\]
Substituting \er{g1sm5} into \er{D1sm1} we obtain \er{D1sm}.

We prove \er{gesm}.
Identities \er{cAsm} and \er{vp0} give
\[
\lb{gsm1}
\begin{aligned}
\l\cA_{12}(x)+\cA_{21}(x)
=\l\vp_2(1-x)\vp_2(x)+\l^2\vp_4(1-x)\vp_4(x)=
z^2\big(g_1(x)-g_2(x)\big),
\\
\cA_{11}(x)+\cA_{22}(x)=\l\vp_4(1-x)\vp_2(x)+\l\vp_2(1-x)\vp_4(x)=
g_1(x)+g_2(x).
\end{aligned}
\]
where
$$
g_1={\cosh z-\cosh z(1-2x)\/4},\qqq
g_2={\cos z-\cos z(1-2x)\/4}.
$$
Thus \er{gsm1} and  \er{fgsm} give
$$
g=\det\ma \vp_2(1)&
z^2(g_1(x)-g_2(x))\\
\vp_4(1)&g_1(x)+g_2(x)\am.
$$
The identities
$$
\vp_2(1)-z^2\vp_4(1)={s_+(1)-s_-(1)\/z}={\sin z\/z},\qqq
\vp_2(1)+z^2\vp_4(1)={s_+(1)+s_-(1)\/z}={\sinh z\/z}
$$
imply \er{gesm}.

We prove \er{D1pim}.
Identity \er{D1sm} gives
$$
D_1((\pi n)^4)={(-1)^n\sinh \pi n\/4\pi n}\int_0^1\g_1(x,(\pi n)^4)
\big(1-\cos 2\pi nx\big)dx.
$$
Using  \er{a1b1sm} we deduce that
$$
D_1((\pi n)^4)={(-1)^n\sinh \pi n\/4\pi n}
\lt({\wh Q_{cn}-\wh Q_0\/(\pi n)^4}-4\int_0^1
\int_0^x\big(\b(s)-\a(s)\big)ds\cos 2\pi nxdx\rt).
$$
Integration by parts yields \er{D1pim}.

ii) The standard arguments based on asymptotics \er{asDsm1}
and Rouch\'e's theorem give the statement.
$\BBox$

\subsection{Proof of Theorem \ref{Thsm}.}
Due to Lemma \ref{LmMMsm},
the function $D(\l,\ve)$, given by \er{Dsm},
is analytic in $(\l,\ve)\in\C\ts\{|\ve|<\ve_1\}$
for some $\ve_1>0$.
Let $|\ve|<\ve_1$. Then, due to Lemma \ref{LmDetsm} ii),
each eigenvalue
$\l_n(\ve), n\ge 1$, is a simple zero of $D(\cdot,\ve)$,
therefore, it is analytic in $\ve$.

We determine eigenvalue asymptotics for the operator $\cE_\ve$ as
$\ve\to 0$.  Let $z_n(\ve)=\l_n^{1\/4}(\ve),n\in\N$. Lemma
\ref{LmDetsm} ii) yields $z_n(\ve)=\pi n+\d$, where $\d=\d(\ve)$
satisfies the estimate $|\d|<1$ for all $\ve$ small enough.
Substituting this asymptotics into \er{asDsm1} we obtain
$$
D(\l_n(\ve))=
{(-1)^n\sinh \pi n\/(\pi n)^2}\d+\ve D_1((\pi n)^4)
+O((|\d|+|\ve|)^2).
$$
The identity $D(\l_n(\ve))=0$ implies $\d=O(\ve)$, which yields
$$
0=D(\l_n(\ve))= {(-1)^n\sinh \pi n\/(\pi n)^2}\d+\ve D_1((\pi n)^4)
+O(\ve^2),
$$
and then
\[
\lb{dsm}
\d= \ve {(-1)^{n+1}(\pi n)^2D_1((\pi n)^4)\/\sinh \pi n}+O(\ve^2).
\]
Substituting \er{D1pim} into \er{dsm} and using
$z_n(\ve)=\pi n+\d$ we deduce that
$$
z_n(\ve)=\pi n+{\ve\/2}(\wh\a_{sn}-\wh\b_{sn})
+{\ve(\wh Q_0-\wh Q_{cn})\/4(\pi n)^3}
+O(\ve^2),
$$
which yields \er{aslnsm}.
$\BBox$

\section{The Barcilon-Gottlieb transformation}
\setcounter{equation}{0}

In this Section we will prove Lemma \ref{lmEuB}.
Identity \er{tx} gives $dt=\x dx$, which yields
\[
\lb{'x't} y_x'=\x y', \qqq \text{here and below}\qqq
y_x'={dy\/dx},\qqq y'={dy\/dt}.
\]
We need the following preliminary results.

\begin{lemma} The operator $\cE$, defined  by
\er{defcE}, satisfies
\[
\lb{EB10}
\cE u={1\/\rho^2}\Big((\rho^2u_{tt}'')_{tt}''
+(2\rho^2\vp u_{t}')_{t}'\Big)+ Q u,
\]
where $\rho$ and  $\vp$ are given by \er{deff} and \er{vp5},
respectively.
\end{lemma}

\no {\bf Proof.}
Identities \er{defcE}, \er{'x't} give
\[
\lb{EBe1}
\cE u={1\/b}(au_{xx}'')_{xx}''+ Q u
={\x\/b}\Big(\x\big(c(\x u')'\big)'\Big)'+ Q u,\qqq
c=a\x.
\]
Moreover, due to $\rho=c^{1\/2}\x$, we get
\[
\lb{xxxxx}
\begin{aligned}
\Big(\x\big(c(\x u')'\big)'\Big)'
=\Big(\x\big(c(\x u''+\x'u')\big)'\Big)'
=\Big(\x\big(c\x u'''+((c\x)'+c\x')u''+(c\x')'u'\big)\Big)'
\\
=\big(\rho^2u'''+(\rho^2)'u''+(c\x')'\x u'\big)'
=(\rho^2u'')''+\big((c\x')'\x u'\big)',
\end{aligned}
\]
and
$$
{(c\x')'\x\/\rho^2}={(c\x')'\/c\x}={\x''\/\x}+{c'\/c}{\x'\/\x}
=\Big({\x'\/\x}\Big)'+\Big({\x'\/\x}\Big)^2+{c'\/c}{\x'\/\x}.
$$
Using the identities
$$
\x'=\e_-,\qqq {c'\/c}={a'\/a}+{\x'\/\x}={4\a+\e_-\/\x},
$$
where $\e_-,\e$ are given by \er{sphi}, we obtain
$$
{(c\x')'\x\/\rho^2}=\Big({\e_-\/\x}\Big)'+\Big({\e_-\/\x}\Big)^2+{(4\a+\e_-)\e_-\/\x^2}
=\Big({\e_-\/\x}\Big)'+{2\e\/\x^2}=2\vp.
$$
Then \er{xxxxx} gives
$$
\Big(\x\big(c(\x u')'\big)'\Big)'
=(\rho^2u'')''+(2\rho^2\vp u')'.
$$
The substitution of this expression   into \er{EBe1} and  the
identity $b=\rho^2\x$ imply \er{EB10}. $\BBox$

\medskip

Identities \er{sigmv}, \er{deff}, \er{sphi} yield
$\rho(x(t))=\rho(0)e^{\int_0^t\s(s)ds}$, then
\[
\lb{fpf}
{\rho'\/\rho}=\s,\qqq
{\rho''\/\rho}=\s^2+\s'.
\]

\begin{lemma}
\lb{Lmbc}
Let
\[
\lb{defy}
y=Uu=\rho u.
\]
Then the function $y$ satisfy \er{4g.dc} iff the function  $u$
satisfy \er{ebdc}.
\end{lemma}

\no {\bf Proof.}
Let $y(0)=y''(0)=0$. Then
$$
u(0)=0,\qqq u_x'(0)=\Big({y\/\rho}\Big)_x'\Big|_{0}
={y_x'\/\rho}\Big|_{0}
$$
and
$$
\begin{aligned}
u_{xx}''(0)=\Big({y\/\rho}\Big)_{xx}''\Big|_{0}
=\Big({y_{xx}''\/\rho}-{2\rho_x'y_x'\/\rho^2}\Big)\Big|_{0}
=\Big({(\x y')'\x\/\rho}-2su_x'\Big)\Big|_{0}
=\Big({\x'\x y'\/\rho}-2su_x'\Big)\Big|_{0}
\\
=(\e_--2s)u_x'\big|_{0}=-2\e_+ u_x'\big|_{0}.
\end{aligned}
$$
Conversely, let $u(0)=(u_{xx}''+2\e_+ u_x')\big|_{0}=0$. Then
$y(0)=0$ and
$$
y''(0)=(2\rho'u'+\rho u'')|_{0}
=\Big({2\rho_x'u_x'\/\x^2}
+{(u_{xx}''-\e_- u_x')\rho\/\x^2}\Big)\Big|_{0}
={\rho\/\x^2}(u_{xx}''+(2s-\e_-)u_x')\big|_{0}=0.
$$
Similar relations at the point $x=1$ hold true, which proves the
statement. $\BBox$

\medskip

\no {\bf Proof of Lemma \ref{lmEuB}.} Let $y=\rho u$. Formula
\er{EB10} implies
\[
\lb{EuB2}
\cE u=\cE {y\/\rho}=
{1\/\rho}\lt({1\/\rho}\lt(\Big(\rho^2\Big({y\/\rho}\Big)''\Big)''
+\Big(2\rho^2\vp\Big({y\/\rho}\Big)'\Big)'\rt)+ Q y\rt).
\]
Using \er{fpf} we obtain
\[
\lb{h1}
{1\/\rho}\Big(\rho^2\vp\Big({y\/\rho}\Big)'\Big)'
={1\/\rho}\big(\rho\vp(y'-\s y)\big)'
=(\vp y')'
-\big(\vp\s^2+(\vp\s)'\big) y.
\]
Moreover,
\[
\lb{f1}
\begin{aligned}
{1\/\rho}\Big(\rho^2\Big({y\/\rho}\Big)''\Big)''
={1\/\rho}\Big(\rho\big(y''-2\s y'+(\s^2-\s')y\big)\Big)''
\\
=y^{(4)}+2\Big(\s^2-{2(\s \rho)'\/\rho}\Big)y''
+{2\/\rho}\Big(\big((\s^2-\s')\rho\big)'-(\s \rho)''\Big)y'
+{((\s^2-\s')\rho)''\/\rho}y
\\
=y^{(4)}+2\lt(\Big(\s^2-{2(\s \rho)'\/\rho}\Big)y'\rt)'
+\Big(2(\s')^2+2\s\s''-\s'''+2(2\s\s'-\s''){\rho'\/\rho}
+(\s^2-\s'){\rho''\/\rho}\Big)y
\\
=y^{(4)}-2\big((2\s'+\s^2)y'\big)'
+\big((\s')^2-\s'''+4\s^2\s'+\s^4\big)y.
\end{aligned}
\]
Substituting \er{h1}, \er{f1}
into \er{EuB2} and using \er{peb}, \er{qeb} we obtain
$$
\cE u={1\/\rho}\big(y^{(4)}+2(py')'+qy\big).
$$
This identity together with Lemma \ref{Lmbc} implies
$$
\cE u={1\/\rho}Hy={1\/\rho}H\rho u,
$$
which yields \er{eqEH}.
$\BBox$

\medskip

\no {\bf Proof of Proposition \ref{propmAmB}.}
Identity \er{qeb} yields
$$
q(t)=\Big(-\s''+{4\/3}\s^3-2\vp\s\Big)'
+(\s')^2+\s^4
-2\vp\s^2+\upsilon.
$$
Substituting this equality  into \er{psi1} we get
$$
\p=\Big(-\s''+{4\/3}\s^3-2\vp\s-{p'\/2}\Big)'
+(\s')^2+\s^4
-2\vp\s^2-{p^2\/2}
+{\wh p_0^2\/2}+\upsilon.
$$
Using  \er{peb}  and the identities
$$
(\s')^2+\s^4-2\vp\s^2-{p^2\/2}=
(\s')^2+\s^4-2\vp\s^2-{(\vp-\s^2-2\s')^2\/2}
={(\s^2-\vp)^2\/2}-(\s'-\vp)^2-{2\/3}(\s^3)',
$$
we obtain
%\er{psimAmB}.
\[
\lb{psimAmB}
\p=\mA_t'+\mB+{\p_0^2\/2}+\upsilon,
\]
where
\[
\lb{mAmB1}
\mA={2\/3}\s^3-2\vp\s+{(\s^2-\vp)_t'\/2},
\qqq
\mB={(\s^2-\vp)^2\/2}-(\s_t'-\vp)^2.
\]
Thus \er{psimAmB} and \er{psi1} yield  \er{defQ}.

We show \er{A2}, \er{B2}. Identity \er{vp5} and
$\x'={\x_x'\/\x}=\e_-$ give
\[
\lb{vp'1} \vp={1\/2}\Big({\e_-\/\x}\Big)'+{\e\/\x^2}
={1\/2}\Big({\e_-'\/\x}-{\e_-^2\/\x^2}\Big)+{\e\/\x^2}, \qq
\vp'={\e_-''\/2\x}-{3\e_-'\e_-\/2\x^2}+{\e_-^3\/\x^3}
+{\e'\/\x^2}-{2\e\e_-\/\x^3}.
\]
 Substituting these identities into \er{mAmB1} and using
$\s={s\/\x}$ we obtain
$$
\mA={2s^3\/3\x^3}-{\e_-^3\/2\x^3}+{\e\e_-\/\x^3}-{2\e s\/\x^3}
+\Big({s-\e_-\/\x}\Big)'{s\/\x}
-{\e_-''\/4\x}+{3\e_-'\e_-\/4\x^2}
-{\e'\/2\x^2}.
$$
Using
$$
\e_-'={(\e_-)_x'\/\x},\qq
\Big({s-\e_-\/\x}\Big)'
={(s-\e_-)_x'-(s-\e_-)\e_-\/\x^2},\qq
\e_-''={(\e_-)_{xx}''-(\e_-)_x'\e_-\/\x^2}
$$
and the formulas  $\e_--2s=-2\e_+$, $\e_-^2-\e=-2\a\e_-$ we get
\er{A2}.

Identities \er{vp5} yield
$$
\begin{aligned}
\s'-\vp=\Big({s-{1\/2}\e_-\/\x}\Big)'
-{\e\/\x^2}=\Big({\e_+\/\x}\Big)'-{\e\/\x^2}
={(\e_+)_x'-2\e\/\x^2},
\\
\s^2-\vp={\vk\/\x^2}-{1\/2}\Big({\e_-\/\x}\Big)'
={2\vk+\e_-^2-(\e_-)_x'\/2\x^2}.
\end{aligned}
$$
Substituting these equalities into \er{mAmB1} we obtain \er{B2}.
$\BBox$

\section{Asymptotics of the determinant $D(\cdot)$}
\setcounter{equation}{0}

\subsection{Preliminaries.}
In this section we determine asymptotics of the determinant $D(\l)$.
Our proof is based on the sharp asymptotics of the monodromy matrix
from our paper \cite{BK4} by using the matrix form of the standard
Birkhoff approach.

Introduce a diagonal $4\ts 4$~-~matrix $\o$  given by
\[
\lb{4g.Om}
\o=\diag(\o_1,\o_2,\o_3,\o_4),\qqq
\o_1=-\o_4=i,\qq\o_2=-\o_3=1,
\]
here and below
$\diag(a_1,a_2,a_3,a_4)=\diag(a_j)_{j=1}^4=(a_k\d_{jk})_{j,k=1}^4$.
Define the set $S^+$ by
$$
\textstyle S^+=\Big\{z\in\C:\arg z\in\big[0,{\pi\/4}\big]\Big\}=
\Big\{\l\in\C_+\Big\},
$$
Note that  we have the following estimates:
\[
\lb{4g.esom}
\Re(i\o_1z)\le\Re(i\o_2z)\le\Re(i\o_3z)\le\Re(i\o_4z)
\qqq\forall\qq z\in S^+.
\]
Introduce a unitary $4\ts 4$~-~matrix $U$ by
\[
\lb{4g.ZQ}
U={1\/2}(\o_k^{j-1})_{j,k=1}^4={1\/2}\ma 1&1&1&1\\
i&1&-1&-i\\
-1&1&1&-1\\
-i&1&-1&i\am.
\]
and a diagonal $4\ts 4$~-~matrix valued function
$$
v(z)=\diag\big(v_j(z)\big)_{j=1}^4,\qqq v_j=v_j(z)=\o_j+{ \wh
p_0\/2z^2}\ol\o_j \qqq\forall\qq j\in\N_4=\{1,2,3,4\},
$$
for $z\in S^+\sm\{0\}$. Note that
\[
\lb{asEE}
e^{iv_jz}=e^{i\o_jz}\big(1+O(|z|^{-1})\big)\qqq\forall\qq j\in\N_4,
\]
as $|z|\to\iy,z\in S^+$.

Define  functions $\vk_{jk}(s,z),j,k\in\N_4,s\in[0,1]$  by
\[
\lb{xijk} \vk_{jk}(s,z)=\ca e^{i(\o_j-\o_k)z(1-s)}
\big(1+{i(\ol\o_j-\ol\o_k)\/2z}\int_s^1p(t)dt\big),&\ j< k
\\
\qqq\qqq\qqq-1\qqq\qqq\qqq\qqq,&\ j=k\\
- e^{-i(\o_j-\o_k)zs}
\big(1-{i(\ol\o_j-\ol\o_k)\/2z}\int_0^sp(t)dt\big),&\ j>k\ac,
\]
and a $4\ts 4$~-~matrix valued function
\[
\lb{4g.cS}
 W(t,z)=\1_4-{p(t)\/z^2}W_1-{p'(t)\/z^3}W_2,
\]
where
\[
\lb{4g.idB}
W_1={1\/4}\ma0&1+i&1-i&1\\-1+i&0&-1&-1-i\\
-1-i&-1&0&-1+i\\ 1&1-i&1+i&0\am,\qq
W_2={1\/8}\ma 0&-2&-2&-1\\2i&0&i&2i\\
-2i&-i&0&-2i\\1&2&2&0\am,
\]
and
\[
\lb{4g.wtQ}
\begin{aligned}
X(t,z)=\big(X_{jk}(t,z)\big)_{j,k=1}^4=p''(t)W_2
+{iq(t)\/4}\ma-i&-i&-i&-i\\-1&-1&-1&-1\\1&1&1&1\\i&i&i&i\am
\\
+{ip^2(t)\/8}\ma i&i&i&0\\ 1&1&0&1
\\-1&0&-1&-1\\0&-i&-i&-i\am
+{3p(t)p'(t)\/16z}\ma-1&1-i&1+i&-{2\/3}\\
1+i&-1&-{2\/3}&1-i\\
1-i&-{2\/3}&-1&1+i\\
-{2\/3}&1+i&1-i&-1\am.
\end{aligned}
\]

Sharp asymptotics of the monodromy matrix $M(1,\l)$ was
determined in \cite{BK4} for the case of
real 1-periodic
coefficients $p,p'',q\in L_\R^1(\R/\Z)$, see
Lemma ~3.1 and relations (3.15), (3.13), (3.8), (3.21), (2.14) in \cite{BK4}.
The proof from \cite{BK4} may be extended
to the case of complex coefficients
$(p,q)\in \mH_2\ts \mH_0^0$ in order to
obtain the following results.

\begin{lemma}
\lb{prL}
Let $(p,q)\in \mH_2\ts \mH_0^0$ and let $r>0$ be large
enough. Then
\[
\lb{4g.MLU}
M(1,\l)=\cZ(z)A(z)e^{i zv(z)}B(z)\cZ^{-1}(z)\qqq
\forall\qq z\in S^+_r=\{z\in S^+:|z|>r\},
\]
where
$$
\cZ(z)=\diag\big(1,iz,(iz)^2,(iz)^3\big),
$$
and $4\ts 4$~-~matrix valued functions $A,B$ are
analytic and uniformly bounded in $S^+_r$.
Moreover,
\[
\lb{4g.LE} A(z)=\big(A_{jk}(z)\big)_{j,k=1}^4=U\Theta(z),\qqq
B(z)=\big(B_{jk}(z)\big)_{j,k=1}^4=\Phi(z)U^*,
\]
\[
\lb{wtAB}
\begin{aligned}
\Theta(z)=\big(\Theta_{jk}(z)\big)_{j,k=1}^4=
W(1,z)\Big(\1_{4}+{\cG(1,z)\/z^3}+{O(1)\/|z|^5}\Big),\\
\Phi(z)=\big(\Phi_{jk}(z)\big)_{j,k=1}^4=
\Big(\1_{4}-{\cG(0,z)\/z^3}+{O(1)\/|z|^5}\Big)W^{-1}(0,z),
\end{aligned}
\]
as $|z|\to\iy$, $z\in S^+$, uniformly on bounded subsets of
$\mH_2\ts \mH_0^0$, where each $4\ts 4$~-~matrix valued function
$\cG(t,z)=\big(\cG_{j k}(t,z)\big)_{j,k=1}^4,t\in\{0,1\},$ is
analytic and uniformly bounded in $S^+_r$ and satisfies
\[
\lb{4g.wtG1a}
\cG_{j k}(1,z)=0\qq\as\qq j\ge k, \qqq \cG_{j k}(0,z)=
0\qq\as\qq j<k,
\]
\[
\lb{4g.wtG1apr}
g_{jk}(z)\ev\ca\cG_{jk}(1,z),\ j<k\\\cG_{jk}(0,z),\
j\ge k\ac =\int_0^1\vk_{jk}(s,z)X_{j k}(s,z)ds.
\]

\end{lemma}

\medskip

Formula \er{4g.MLU} represents the monodromy matrix as a  product of
the simple diagonal matrices $\cZ,\cZ^{-1}$, the bounded matrices
$A,B$, and the diagonal exponential factor $e^{izv}$. The following
result is an immediate consequence of identity \er{4g.MLU}.

\begin{lemma}
Let $(p,q)\in \mH_2\ts \mH_0^0$. Then the function $D(\l)$, given by
\er{defD}, satisfies
\[
\lb{detPhi}
D(\l)=-{1\/z^2}\sum_{j,k\in\N_4\atop j<k}e^{i(v_j+v_k)z}
\det\big(\a_{jk}(z)\b_{jk}(z)\big),\qqq \forall \ z\in S^+_r,
\]
for some $r>0$ large enough.  Here the $2\ts 2$~-~matrix valued
functions $\a_{jk},\b_{jk}$ have the form
\[
\lb{4g.MaMb}
\a_{jk}=\ma A_{1j}&A_{1k}\\A_{3j}&A_{3k}\am,\qqq
\b_{jk}=\ma B_{j2}&B_{j4}\\B_{k2}&B_{k4}\am,\qqq j,k\in\N_4,
\]
and $A_{jk},B_{jk}$ are given by \er{4g.LE}.
\end{lemma}

\no {\bf Proof.} Identity \er{4g.MLU} gives
$$
\ma M_{12}&M_{14}\\M_{32}&M_{34}\am(1,\l)
=\ma1&0\\0&(iz)^2\am\sum_{j=1}^4e^{iv_jz}\g_{jj}(z)\ma(iz)^{-1}&0\\0&(iz)^{-3}\am,
$$
where $ \g_{jk}=\ma
A_{1j}B_{j2}&A_{1k}B_{k4}\\A_{3j}B_{j2}&A_{3k}B_{k4}\am. $ Then
$$
D(\l)=\det\ma M_{12}&M_{14}\\M_{32}&M_{34}\am(1,\l)
=-{1\/z^2}\det\sum_{j=1}^4e^{iv_jz}\g_{jj}(z).
$$
Using the identities
$$
\det\sum_{j=1}^4e^{iv_jz}\g_{jj}
=\sum_{j,k=1}^4e^{i(v_j+v_k)z}\det \g_{jk} =
\sum_{1\le j<k\le 4}e^{i(v_j+v_k)z} (\det \g_{jk}+\det \g_{kj})
$$
and $ \det \g_{jk}+\det \g_{kj}=\det(\a_{jk}\b_{jk}) $ we obtain
\er{detPhi}. $\BBox$

\subsection{Rough asymptotics of $D$.}
Now we determine rough asymptotics of the determinant.

\no {\bf Proof of Lemma \ref{LmrAsF}.} Let $|z|\to\iy,z\in S^+$.
Substituting \er{4g.cS} into \er{wtAB}  we obtain
$$
\Theta(z)=\1_4+O(|z|^{-1}),\qqq \Phi(z)=\1_4+O(|z|^{-1}).
$$
Then identities \er{4g.LE} imply
$$
A(z)=U+O(|z|^{-1}),\qqq B(z)=U^*+O(|z|^{-1}),
$$
which jointly with \er{detPhi} yield
\[
\lb{asD1} D(\l)=-{1\/z^2}\sum_{j,k\in\N_4\atop j<k}e^{i(v_j+v_k)z}
\det\lt( \ma U_{1j}&U_{1k}\\U_{3j}&U_{3k}\am \ma \ol U_{2j}&\ol
U_{4j}\\\ol U_{2k}&\ol U_{4k}\am +{O(1)\/|z|}\rt).
\]
Definition \er{4g.ZQ} gives
$$
\det \ma U_{1j}&U_{1k}\\U_{3j}&U_{3k}\am \ma \ol U_{2j}&\ol
U_{4j}\\\ol U_{2k}&\ol U_{4k}\am ={1\/16}\det \ma
1&1\\\o_{j}^2&\o_{k}^2\am \ma \ol \o_{j}&\ol \o_{j}^3\\\ol
\o_{k}&\ol \o_{k}^3\am
={1\/8}\Big(\ol\o_j\ol\o_k-{1\/\ol\o_j\ol\o_k}\Big),
$$
and using \er{asD1} we have
$$
D(\l)=-{1\/8z^2}\sum_{j,k\in\N_4\atop j<k}e^{i(v_j+v_k)z}
\Big(\ol\o_j\ol\o_k-{1\/\ol\o_j\ol\o_k}+{O(1)\/|z|}\Big).
$$
Then due to \er{4g.Om}, \er{asEE}  we obtain
$$
D(\l)
=-{i\/2z^2}\Big(\cosh \big((1+i)z\big)\big(1+{O(1)\/|z|}\big)
-\cosh\big((1-i)z\big)\big(1+{O(1)\/|z|}\big)\Big),
$$
which yields \er{asF5} in $S^+$. By the identity $D(\ol\l,\ol p,\ol
q)=\ol{D}(\l,p,q)$, we extend \er{asF5} from $S^+$ onto $S$. $\BBox$

\subsection{Sharp asymptotics of $D$.}
In order to obtain asymptotics \er{4g.adwM12} we have to improve asymptotics
\er{asF5}. Substituting
\er{asEE} into \er{detPhi} and using \er{4g.Om} we obtain
\[
\lb{decab}
D(\l)=-{e^{iv_4z}\/z^2}
\Big(e^{iv_2z}d_2(z)+e^{iv_3z}d_3(z)+O(e^{-\Re z})\Big)
\]
as $|z|\to\iy,z\in S^+$, uniformly on bounded subsets of $\mH_2\ts
\mH_0^0$, where
\[
\lb{defdj}
d_j(z)=\det\big(\a_{j4}(z)\b_{j4}(z)\big),\qqq j=2,3.
\]
Below we will need the following auxiliary identities.

\begin{lemma}  Let $(p,q)\in \mH_2\ts \mH_0^0$.
The functions $d_j,j=2,3$, satisfy the following identities
\[
\lb{idal}
d_j=-{i\/4}\det(\xi_j\eta_j),
\]
where
\[
\lb{mAmB}
\xi_j=\ma\Theta_{14}+\Theta_{44}&\Theta_{1j}+\Theta_{4j}\\
\Theta_{24}+\Theta_{34}&\Theta_{2j}+\Theta_{3j}\am,\qqq
\eta_j=\ma\Phi_{j2}-\Phi_{j3}&\Phi_{j1}-\Phi_{j4}\\
\Phi_{42}-\Phi_{43}&\Phi_{41}-\Phi_{44}\am,
\]
and $\Theta_{jk},\Phi_{jk}$ are given by \er{wtAB}.
\end{lemma}

\no {\bf Proof.} Let $j=2,3$. Identities \er{4g.LE}, \er{4g.MaMb}
provide
\[
\lb{cBcAj4} \a_{j4}=u_1\theta_{1j}+u_2\theta_{2j},\qqq
\b_{j4}=\phi_{j1}u_3+\phi_{j2}u_4,
\]
where
\[
\lb{zd4}
\theta_{1j}=\ma\Theta_{1j}&\Theta_{14}\\\Theta_{4j}&\Theta_{44}\am,
\qq
\theta_{2j}=\ma\Theta_{2j}&\Theta_{24}\\
\Theta_{3j}&\Theta_{34}\am, \qq
\phi_{j1}=\ma\Phi_{j1}&\Phi_{j4}\\
\Phi_{41}&\Phi_{44}\am, \qq
\phi_{j2}=\ma\Phi_{j2}&\Phi_{j3}\\
\Phi_{42}&\Phi_{43}\am,
\]
$$
u_1=\ma U_{11}&U_{14}\\
U_{31}&U_{34}\am,\qq
u_2=\ma U_{12}&U_{13}\\
U_{32}&U_{33}\am, \qq u_3=\ma \ol U_{21}&\ol U_{41}\\\ol U_{24}&\ol
U_{44}\am,\qq
u_4=\ma \ol U_{22}&\ol U_{42}\\
\ol U_{23}&\ol U_{43}\am.
$$
Definition \er{4g.ZQ} give
$$
u_2={1\/2}\ma 1&1\\1&1\am,\qq u_3={i\/2}\ma -1&1\\1&-1\am,\qq
u_1=u_4={J\/2},\qq\where\qq J=\ma 1&1\\-1&-1\am,
$$
and we have
$$
u_3u_2=u_4u_1=0,\qqq iu_3u_1=u_4u_2={J\/2}.
$$
Then identities \er{cBcAj4} imply
$$
\begin{aligned}
\b_{j4}\a_{j4}
={1\/2}(\phi_{j2}J\theta_{2j}-i\phi_{j1}J\theta_{1j}).
\end{aligned}
$$
Substituting \er{zd4} into this identity we obtain
$$
\b_{j4}\a_{j4}
={1\/2}\ma \Phi_{j1}-\Phi_{j4}&\Phi_{j2}-\Phi_{j3}\\
\Phi_{41}-\Phi_{44}&\Phi_{42}-\Phi_{43}\am \ma-i&0\\0&1\am\ma
\Theta_{1j}+\Theta_{4j}&\Theta_{14}+\Theta_{44}\\
\Theta_{2j}+\Theta_{3j}&\Theta_{24}+\Theta_{34}\am.
$$
Identity \er{defdj} gives \er{idal}. $\BBox$

%%%%%%%%%%%%%%%%%%%
\medskip

We will determine asymptotics of the functions $d_j(z)$.

\begin{lemma} Let $(p,q)\in \mH_2\ts \mH_0^0$.

i) The functions $d_j(z)$ satisfy the asymptotics
\[
\lb{asdeta}
\begin{aligned}
d_2(z)
={i\/4}\Big(1+{1+i\/8z^3}\big(p'(1)-p'(0)\big)-{g_{22}(z)-f_2(z)\/z^3}
\Big) +{O(1)\/|z|^5},
\\
d_3(z)
=-{i\/4}\Big(1+{1-i\/8z^3}\big(p'(1)-p'(0)\big)+{f_1(z)+f_2(z)\/z^3}
\Big) +{O(1)\/|z|^5},
\end{aligned}
\]
as $|z|\to\iy,z\in S^+$,  uniformly on bounded subsets of $\mH_2\ts
\mH_0^0$, where
\[
\lb{ab} f_1=g_{23}+g_{32}-g_{33},\qq
f_2=g_{41}+g_{14}-g_{44}+{3(p^2(1)-p^2(0))\/16z},
\]
and $g_{jk}$ are given by \er{4g.wtG1apr}.

ii) Let $z\in S^+_r$ for some $r>0$ large enough. Then
\[
\lb{4g.wtg2i}
g_{jj}(z)=-{i\o_jP_2\/8}
+{3\big(p^2(1)-p^2(0)\big)\/32z}\qqq\forall\qq j\in\N_4,\qqq P_2=\int_0^1p^2(t)dt.
\]
\end{lemma}

\no {\bf Proof.} i) Let $|z|\to\iy$ in $S^+$.  Identities
\er{4g.cS}, \er{wtAB} and $W_1^2=-{3\/16}\1_4$ imply
\[
\lb{asLL-} \Theta(z)=\1_4-{p(1)\/z^2}W_1
+{1\/z^3}\big(-p'(1)W_2+\cG(1,z)\big)+{O(1)\/|z|^5},
\]
\[
\lb{asLL-1} \Phi(z)=\1_4+{p(0)\/z^2}W_1
+{1\/z^3}\Big(p'(0)W_2-\cG(0,z)-{3p^2(0)\/16z}\1_4\Big)+{O(1)\/|z|^5}.
\]

Substituting identities \er{4g.idB}, \er{4g.wtG1a} into asymptotics
\er{asLL-}, \er{asLL-1} and using \er{mAmB} we obtain
$$
\xi_2(z)=X_1 +{p'(1)\/8z^3}X_2 +{1\/z^3}\ma
g_{14}&g_{12}\\g_{24}+g_{34}&0 \am(z)+{O(1)\/|z|^5},
$$
$$
\xi_3(z)=X_1 +{p'(1)\/8z^3}X_2^* +{1\/z^3}\ma
g_{14}&g_{13}\\g_{24}+g_{34}&g_{23}\am(z)+{O(1)\/|z|^5},
$$
$$
\eta_2(z)=\ma1&0 \\0&-1\am\lt(Y_1 +{p'(0)\/8z^3}Y_2
-{1\/z^3}\ma g_{22}&g_{21}\\
g_{43}-g_{42}&g_{44}-g_{41}\am(z)\rt)+{O(1)\/|z|^5},
$$
$$
\eta_3(z)=-\lt(Y_1 +{p'(0)\/8z^3}Y_2^* -{1\/z^3}\ma
g_{33}-g_{32}&-g_{31}\\g_{43}-g_{42}&g_{44}-g_{41}\am(z)
\rt)+{O(1)\/|z|^5},
$$
where
$$
\begin{aligned}
X_1=\1_2 -{p(1)\/4z^2}\ma1&2\\ -2&-1\am,\qqq
X_2=\ma1&0 \\0&i\am,\\
Y_1=\1_2 +{p(0)\/4z^2}\ma1&2i\\2i&-1\am-{3p^2(0)\/16z^4}\1_2,\qqq
Y_2=\ma-i&0\\0&-1\am.
\end{aligned}
$$
The standard formula for any $n\ts n$-matrix $\cA$
$$
\det(\1_n+\ve\cA)=1+\ve\Tr\cA+
{\ve^2\/2}\Big((\Tr\cA)^2-\Tr\cA^2\Big) +O(\ve^3)\qqq\as\qq|\ve|\to
0,\qq\ve\in\C,
$$
yields
$$
\det\xi_2(z)=1+x_1(1)+{g_{14}(z)\/z^3}
+{O(1)\/|z|^5},
\qq
\det\xi_3(z)=1+x_2(1)
+{g_{23}(z)+g_{14}(z)\/z^3}+{O(1)\/|z|^5},
$$
$$
\det\eta_2(z)=-1+x_1(0)-{g_{41}(z)-g_{44}(z)-g_{22}(z)\/z^3}
+{O(1)\/|z|^5},
$$
$$
\det\eta_3(z)=1-x_2(0)
+{g_{41}(z)-g_{44}(z)+g_{32}(z)-g_{33}(z)\/z^3}+{O(1)\/|z|^5}.
$$
where
$$
x_1(t)={(1+i)p'(t)\/8z^3}+{3p^2(t)\/16z^4},\qqq
x_2(t)={(1-i)p'(t)\/8z^3}+{3p^2(t)\/16z^4}
$$
Substituting these asymptotics into \er{idal} we obtain
\er{asdeta}.

ii) Identities \er{xijk} yield $\vk_{jj}=-1$, which together
with  \er{4g.wtG1apr} give  $g_{jj}=-\int_0^1 X_{jj}ds$.
Substituting \er{4g.wtQ} into the last identity and using
$\int_0^1q(t)dt=0$ we obtain \er{4g.wtg2i}. $\BBox$

\medskip

We will determine the sharp asymptotics of the determinant.

\no {\bf Proof of Lemma \ref{Lmascf} i), iii).} i) Let
$|z|\to\iy,z\in S^+$. Asymptotics \er{asdeta} and \er{decab} imply
\er{4g.adwM12}, where
\[
\lb{aswtF}
F(z)={\big(p'(1)-p'(0)\big)\sin v_2z\/8}
-\vp_1(z)+f_2(z)\sin v_2z
+{O(e^{\Im z})\/|z|^2},
\]
\[
\lb{A(z)}
\vp_1(z)={g_{22}(z)e^{iv_2z}+f_1(z)e^{-iv_2z}\/2i}.
\]
Asymptotics \er{aswtF} yields \er{estF}.
Using the identity $D(\ol\l,\ol p,\ol q)=\ol{D}(\l,p,q)$
we extend \er{4g.adwM12}, \er{estF} from $S^+$ onto $S$.

iii) Let  $|z|\to\iy,z\in S^+$.  Substituting identities
\er{xijk}, \er{4g.wtQ} into \er{4g.wtG1apr},
using \er{4g.Om} and integrating by parts we
obtain
\[
\lb{asG2332a}
\begin{aligned}
g_{23}(z)= -{i\/4}\int_0^1
e^{i2z(1-s)}\Big(1+{i\/z}\int_s^1p(t)dt\Big)V(s) ds+{O(1)\/|z|^2},
\\
g_{32}(z)= -{i\/4}\int_0^1 e^{i2zs}
\Big(1+{i\/z}\int_0^sp(t)dt\Big)V(s)ds+{O(1)\/|z|^2}.
\end{aligned}
\]
Asymptotics   \er{asG2332a}, \er{4g.wtg2i},   \er{ab} and \er{A(z)}
yield
\[
\lb{A1}
\vp_1(z)=-{P_2\/8}\cos v_2z
+{3(p^2(1)-p^2(0))\/32 z}\sin v_2z
-{1\/4}\int_0^1\cos z(1-2s)V(s)ds+\wt\vp_1(z),
\]
where
\[
\lb{aswtal}
\wt\vp_1(z)=\ca o(e^{\Im z}|z|^{-1}),j=0\\O(e^{\Im z}|z|^{-2}),j=1\ac.
\]
Similar arguments give
$$
\begin{aligned}
g_{14}(z)= -{1\/4}\int_0^1
e^{-2z(1-s)}\Big(1+{1\/z}\int_s^1p(t)dt\Big)V(s) ds+{O(1)\/|z|^2},
\\
g_{41}(z)= -{1\/4}\int_0^1 e^{-2zs}
\Big(1+{1\/z}\int_0^sp(t)dt\Big)V(s)ds+{O(1)\/|z|^2},
\end{aligned}
$$
which yields
\[
\lb{b(z)1}
f_2(z)={P_2\/8}-{e^{-z}\/2}\int_0^1\cosh z(1-2s)V(s)ds+\wt f_2(z),
\]
where
\[
\lb{aswtbe}
\wt f_2(z)=\ca o(|z|^{-1}),j=0\\
O(|z|^{-2}),j=1\ac.
\]
Substituting \er{A1}, \er{b(z)1} into \er{aswtF}
we obtain
\[
\lb{asF111}
F(z)={P_2\/8}\cos v_2z+{1\/4}\int_0^1\cos z(1-2s)V(s)ds
+\vp_2(z)\sin v_2z+\wt F(z),
\]
where
$$
\vp_2(z)={\big(p'(1)-p'(0)\big)\/8}-{3(p^2(1)-p^2(0))\/32 z}
+{P_2\/8}-{e^{-z}\/2}\int_0^1\cosh z(1-2s)V(s)ds,
$$
$$
\wt F(z)=\wt\vp_1(z)+\wt f_2(z)\sin v_2z
+{O(e^{\Im z})\/|z|^2}.
$$

Asymptotics \er{aswtal}, \er{aswtbe} give
\[
\lb{zd5}
\wt F(z)=\ca o(e^{\Im z}|z|^{-1}),j=0\\
O(e^{\Im z}|z|^{-2}),j=1\ac
\]
as $|z|\to\iy,z\in S^+$.
Using the identity $D(\ol\l,\ol p,\ol q)=\ol{D}(\l,p,q)$
we extend identity \er{asF111} and asymptotics \er{zd5} from $S^+$ onto $S$.
Relations  \er{asF111}, \er{zd5} give  \er{aswF1}, \er{awF1}.
$\BBox$

\medskip

\setlength{\itemsep}{-\parskip} \footnotesize \no {\bf
Acknowledgments.} { Evgeny Korotyaev's  study was partly supported
by the RFFI grant  No 11-01-00458 and by  project  SPbGU No
11.38.215.2014.}

%%%%%%%%%%%%%%%%%%%%%%%%%%%%%%%%END

\end{document}